\documentclass[a4paper,fleqn]{cas-sc}
\usepackage{algorithm} 
\usepackage{algorithmic}
\usepackage{subfigure}

\usepackage[authoryear,longnamesfirst]{natbib}

\newcommand{\PreserveBackslash}[1]{\let\temp=\\#1\let\\=\temp}
\newcolumntype{C}[1]{>{\PreserveBackslash\centering}p{#1}}
\newcolumntype{R}[1]{>{\PreserveBackslash\raggedleft}p{#1}}
\newcolumntype{L}[1]{>{\PreserveBackslash\raggedright}p{#1}}
\def\tsc#1{\csdef{#1}{\textsc{\lowercase{#1}}\xspace}}
\tsc{WGM}
\tsc{QE}

\begin{document}
\let\WriteBookmarks\relax
\def\floatpagepagefraction{1}
\def\textpagefraction{.001}

\shorttitle{}
\shortauthors{Shuo Yu et al.}

\title [mode = title]{Collaborative Team Recognition: A Core Plus Extension Structure}

%

%

\author[1]{Shuo Yu}
\affiliation[1]{organization={School of Computer Science and Technology},
	addressline={Dalian University of Technology},
	state={Dalian 116024},
	country={China}}
\author[2]{  Fayez Alqahtani}
\affiliation[2]{organization={Software Engineering Department, College of Computer and Information Sciences},
	addressline={King Saud University},
	state={Riyadh 12372},
	country={Saudi Arabia}}
\author[3]{  Amr Tolba}
\affiliation[3]{organization={Computer Science Department, Community College},
	addressline={King Saud University},
	state={Riyadh 11437},
	country={Saudi Arabia}}
\author[4]{  Ivan Lee}
\affiliation[4]{organization={STEM},
	addressline={University of South Australia},
	city={Adelaide},
	state={SA 5001},
	country={Australia}}
\author[5]{  Tao Jia}
\affiliation[5]{organization={College of Computer and Information Science},
	addressline={Southwest University},
	state={Chongqing 400715},
	country={China}}
\author[6]{  Feng Xia\corref{cor1}}
\ead{f.xia@ieee.org }
\affiliation[6]{organization={Institute of Innovation, Science and Sustainability},
	addressline={Federation University Australia},
	city={Ballarat},
	state={VIC 3353},
	country={Australia}}
\cortext[cor1]{Corresponding author}


\begin{abstract}
	Scientific collaboration is a significant behavior in knowledge creation and idea exchange. To tackle large and complex research questions, a trend of team formation has been observed in recent decades. In this study, we focus on recognizing collaborative teams and exploring inner patterns using scholarly big graph data. We propose a collaborative team recognition (CORE) model with a "core + extension" team structure to recognize collaborative teams in large academic networks. In CORE, we combine an effective evaluation index called the collaboration intensity index with a series of structural features to recognize collaborative teams in which members are in close collaboration relationships. Then, CORE is used to guide the core team members to their extension members. CORE can also serve as the foundation for team-based research. The simulation results indicate that CORE reveals inner patterns of scientific collaboration: senior scholars have broad collaborative relationships and fixed collaboration patterns, which are the underlying mechanisms of team assembly. The experimental results demonstrate that CORE is promising compared with state-of-the-art methods.
\end{abstract}


\begin{highlights}
\item A fine-grained collaborative team recognition method is proposed.
\item Collaborative teams are formulated with "core+extension" structure.
\item The underlying relationship between team output and collaboration intensity is found.
\item It is found that core members have broad collaboration relationships and fixed collaboration patterns.
\end{highlights}


\maketitle

\section{Introduction}
{S}cholarly big graph data brings both opportunities and challenges for big graph data management in exploring scientific collaboration patterns ~\citep{xia2017big,zhou2021academic}. A new research field called science of science has emerged to investigate the theoretical basis of scholars' behaviors from various perspectives ~\citep{yu2021fa,fortunato2018science,yu2019science,jia2017quantifying,DBLP:journals/joi/TothLVJPFE21}. Growing interest has been observed in identifying fundamental drivers of science and promoting scientific research, and studies have shown that great breakthroughs are often associated with exceptional teams~\citep{wu2019large}. This phenomenon has also been observed in the evolution of scientific collaborations over the past century, indicating that collaborative teams, rather than individual scientists, have become fundamental units for scientific discovery~\citep{mao2019data}.
	
As the scientific knowledge frontier expands, relationships embodied in scholarly endeavors, such as research papers, patents, and books, provide much information about social structures and knowledge dissemination among scholars~\citep{xie2014undemocracy,wang2018acekg,DBLP:journals/scientometrics/JinYSHT21,DBLP:journals/tkdd/WangXWGTD21}. Collaboration has become a common practice, as reflected in the increasing number of co-authored papers and the growing size of scientific teams~\citep{wuchty2007increasing,yu2019science,2018Assessing}. According to previous studies~\citep{dong2017century,zhu2021101205}, team-based research dominates scientific production. Therefore, many research teams have been assembled to explore novel ideas and promote creativity. This phenomenon has inspired the investigation of team formulation and scientific achievements. For example, several studies investigated the structure, formation, and evolution of scientific teams and their influences on team outcomes and diversity ~\citep{national2015enhancing,coccia2016evolution,yang2016forming,smith2021101104}. Some studies focused on recommending appropriate collaborators to scholars based on their collaborative patterns ~\citep{li2017enhancing,yu2019academic,zhang2019feature}.
	
	Various fields benefit from team formation and collaboration patterns ~\citep{milojevic2014principles,zhang2018understanding,DBLP:journals/jasis/YaoZQT20}. First, mining knowledge from collaboration patterns provides scholars with further insights into science and enhances the management of scholarly big data. Second, a comprehensive evaluation of team performance can be established by analyzing collaborative behaviors. Research on teams and collaborations has a long history across various disciplines. Over the last few decades, many sociologists and managerialists have investigated teams in communication networks (e.g., mobile and email networks)~\citep{yuan2020international,shao2020ai,xu2020multivariate} and business trading networks~\citep{DBLP:journals/ijon/WangHG20}. In this era of big data, scholars have investigated teams from a new perspective. Many computer science studies have systematically addressed this issue by analyzing large-scale networks~\citep{xia2021chief,wang2017scientific,cen2019representation,qu2017an,bedru2020big,DBLP:journals/tbd/LinTTC17}. Collaborations among scholars can be constructed as academic social networks, such as keyword, co-author, and citation networks. Therefore, scholars can use several advanced theories to investigate the mechanisms and evolution of scientific collaboration~\citep{newman2001structure,azoulay2018toward,zhang2020data}.
	
	The concept of \textbf{teams} is qualitatively defined from various perspectives. Guzzo~\textit{et al.} defined teams as groups of individuals who are seen as social entities by others and who are interdependent with the tasks they perform~\citep{guzzo1996teams}. Huitt~\textit{et al.} considered teams to be project or research groups that work together ~\citep{huitt2015team}. Milojevi{\'{c}} defined collaborative teams as researchers who are co-authors of their research articles, which can easily quantify the manifestations of team effort ~\citep{milojevic2014principles}. Scientific teams have special characteristics compared with other organizational systems; therefore, we cannot simply define scholars who co-authored papers or who work in the same institutions as collaborative teams. However, the lack of real datasets poses a challenge for related research. In addition, collaboration patterns are generally dynamic, which requires significant effort to update the datasets. Exploring team properties using large-scale article datasets can yield accurate and meaningful conclusions.
	
	In this study, we explicitly determine how collaborative teams are quantitatively organized based on large-scale academic graph data. We define the collaborative team as a group of scholars with the following properties: (1) scholars are self-assembled, (2) scholars maintain close collaborative relationships with each other, and (3) scholars are interdependent and interactive in terms of mutual interests, such as papers, patents, and books. We propose \textbf{CORE} (collaborative team recognition) to recognize collaborative teams using the aforementioned properties. A collaborative team is composed of two parts, that is, a core team and extension team; leaders are regarded as part of the core team. This structure regards a few members as cores who are highly interconnected, and the rest of the members as extensions who attach to the cores.
	
	We apply the collaboration intensity index (CII)~\citep{yu2017team} to measure the collaboration intensity between scholars and to recognize nonoverlapping academic teams. This study extends prior research by adopting network science theories and scholars' academic characteristics to evaluate collaborative behaviors synthetically according to a co-author network weighted by the CII. With the leadership qualifications of core members, the proposed CORE method further enhances the recognition algorithm. It was found that academic teams generally contain core teams and extended groups. Based on this discovered pattern in academic teams, we propose an academic team recognition algorithm, CORE, which can effectively identify overlapping academic teams. Furthermore, we employ a large-scale scholarly dataset, that is, Microsoft Academic Graph-computer science discipline (MAG-CS), to comprehensively explore the collaboration patterns of scholars. MAG-CS contains entire information about authors and papers. By analyzing collaborative teams, some collaboration patterns of scholars in science are captured. In summary, the contributions of this paper are summarized as follows:
	
	\begin{itemize}
		\item \textbf{Core plus extended academic team structure definition:} we define collaborative team structure according to the collaboration intensity between scholars, which are composed of two parts, i.e., core teams and extension groups. core teams contain members with higher academic reputation but lower CII, while extension groups contain members with opposite attributes.
		\item \textbf{Academic team recognition based on CII:} based on this fundamental structure of academic team, CORE is then proposed to recognize collaborative teams in academia. The core teams are firstly recognized according to limited academic attributes, and the extension members are then recognized by qualifying collaboration influence of core team members.
		\item \textbf{Collaborative pattern analytics in large-scale academic data set:} we recognize academic teams with proposed CORE in MAG-CS, which is a large-scale data set to better manage scholarly big graph data. The collaboration patterns uncover the team assembling the mechanisms and evolution of collaboration networks are respectively analyzed in different periods.
	\end{itemize}
	
The paper is organized as follows. Section~\ref{sec2} introduces the definition of collaborative team recognition problem and some scientific evaluation metrics. Section~\ref{sec3} illustrates our proposed model CORE. Section~\ref{sec4} evaluates the performance of CORE as well as other state-of-the-arts methods in MAG-CS. Section~\ref{sec5} analyzes and discusses the experimental results. Section~\ref{sec6} concludes the paper.

\section{Preliminaries}~\label{sec2}

In this section, we formally introduce five relevant evaluation metrics and describe the problem of collaborative team recognition.

\subsection{Scientific Evaluation Metrics}
With the advancement of contemporary science, an increasing number of studies have focused on evaluating scholars and measuring their scientific collaborations from different perspectives. In the following section, we review five metrics to analyze collaboration patterns from different perspectives.
	
\subsubsection{Collaborative Frequency}
Collaborative frequency is a major attribute of collaboration, which can be quantified as the number of papers co-authored by two scholars. It is generally believed that the higher the collaborative frequency between two scholars, the closer their collaboration.
	
\subsubsection{Partnership Ability Index}
The partnership ability index (PHI) was proposed ~\citep{schubert2011hirsch} to combine the number of collaborators and collaborative frequency of a given scholar. In co-author networks, the value of the PHI for a scholar is defined as follows:
\textit{A scholar has PHI $\varphi$, if they collaborate with $\varphi$ of their n collaborators, and the scholar has at least $\varphi$ co-authored papers with each collaborator. For the other (n - $\varphi$) collaborators, the scholar has no more than $\varphi$ papers co-authored with each collaborator.}
For example, PHI for a scholar 20 denotes that he/she has collaborated with 20 different scholars at least, and has published at least 20 papers with each collaborator. The PHI reflects a scholar's embedding situation in their own collaboration network. A low PHI value indicates that a scholar maintains loose or fragile relationships with others. By contrast, high PHI values reflect broad, sustained, and stable collaborative relationships with others.
	
\subsubsection{Collaboration Intensity Index}
The CII considers the collaborative frequency between two scholars and the number of publications in the year interval ~\citep{yu2017team}. The CII between scholars \textit{i} and \textit{j} from $ y_{1} $ to $ y_{2} $ is calculated as follows:
\begin{equation}
	CII=\frac{\bigtriangleup k_{ij}^{2}}{\bigtriangleup k_{i}\bigtriangleup k_{j}},
\end{equation}
where $ \bigtriangleup k_{i} $ represents the number of published papers by scholar $ i $ from time $ y_{1} $ to $ y_{2} $ and $ \bigtriangleup k_{j} $ is the counterpart of scholar $ j $; $ \bigtriangleup k_{ij} $ is the number of co-authored papers by scholars $ i $ and $j $ from time $ y_{1} $ to $ y_{2} $. When two scholars have a higher CII value, it means that they have a closer relationship with each other. Particularly, the value of CII reaches 1 if two scholars collaborate with each other on all papers. A lower CII indicates a looser relationship between scholars. Compared with basic metrics, the CII is a relative and efficient index for measuring collaboration relationships.
	
There are also some metrics to evaluate scholars not only by their collaboration behaviors, but also by their individual academic behaviors. For example, scholars may have different behaviors during their different stages in academic careers. Thus, we adopt the following two metrics to find professional scholars who are likely to be core members of teams.
	
\subsubsection{Academic Age}
The relationships between individual academic careers and scientific achievements are related to one's own collaboration patterns ~\citep{liu2018hot}. Academic age, which is defined as the year intervals between the given year and the year the scholar published their first paper, provides a method to analyze scholars' scientific careers. Academic careers can be divided into different stages according to the academic age of the scholars ~\citep{LU2021101207}. Based on research by Wang \textit{et al.}, a scholar's academic career can be divided into three stages: beginner (early career), junior (rising career), and senior (mature career) ~\citep{wang2017scientific}. It provides a new perspective to explore core members of collaborative teams according to academic age, which reflects scholars' productivity and reputations.
	
\subsubsection{Academic Vitality}
Academic vitality is a basic metric used to measure a scholar's productivity, which is defined as the average number of the scholar's annual publications over a certain duration. The academic vitality of scholar $ i $ can be computed as:
\begin{equation}
	AV(i) = \frac{\Delta N_{i}}{\Delta T},
\end{equation}
where $ {\Delta N_{i}} $ is the number of papers published by scholar $ i $ from $ y_{1} $ to $ y_{2} $, and $ \Delta T = |y_{2}-y_{1}| $ represents the year intervals between $ y_{1} $ and $ y_{2} $. Academic vitality reflects an individual's activities during this time interval. A higher value of academic vitality indicates that a scholar is more productive and active. The main purpose of using this index is to measure the activities of scholars. Scholars with active productivity at a given time often have frequent collaboration behaviors, which can help recognize teams accurately and understand collaboration patterns comprehensively.

\subsection{Problem Definition}
To recognize collaborative teams, we first modelled the academic network. An undirected graph is denoted by \textit{G= ($ V_w, E_w $)}, where \textit{$ V_w $} is a weighted vertex set and \textit{$ E_w $} is a weighted edge set. In the graph, we connect the entities represented by nodes through the relationships between the entities. In other words, if two nodes are connected, there must be some connection between them. We consider scholars as nodes and the collaborative relationships between scholars as edge weights in the academic network. Obviously, the co-authorship graph is a 1-mode graph. After constructing the academic network from the data, we extracted clusters with the structure of "core + extension" by their academic age and academic vitality. The collaborative team recognition problem aims to find the set of overlapping clusters \textit{C} from \textit{G}, which ensures that the clusters in \textit{C} are of the structure of "core + extension". In other words, we aim to find the resulting cluster set $ \psi = \{C_1, ..., C_n\} $, whose elements are subgraphs with collections of nodes and edges between them. Moreover, $ C_1 \cup ...\cup C_n \subsetneqq V_w $ and $ {\exists}_{1 \leq {i < j} \leq {n}}\  C_i \cap {C_j} \neq \varnothing $. $C_i$ comprises two sets: core set $Core_i$ and extension set $Extension_i$. $Core_i$=$\{AA(i)>\alpha\cap{AV(i)>\beta}\}$, where $AA(i)$ refers to the academic age of scholar $i$ and $AV(i)$ refers to the academic vitality of scholar $i$.

\section{Design of CORE}~\label{sec3}
	In this section, we propose CORE, which is an efficient model that delivers a feasible solution to the problem of collaborative team recognition. Then, we analyze the complexity of the proposed method.
\begin{figure}[htpb]
	\centering
	\includegraphics[width=1.0\textwidth]{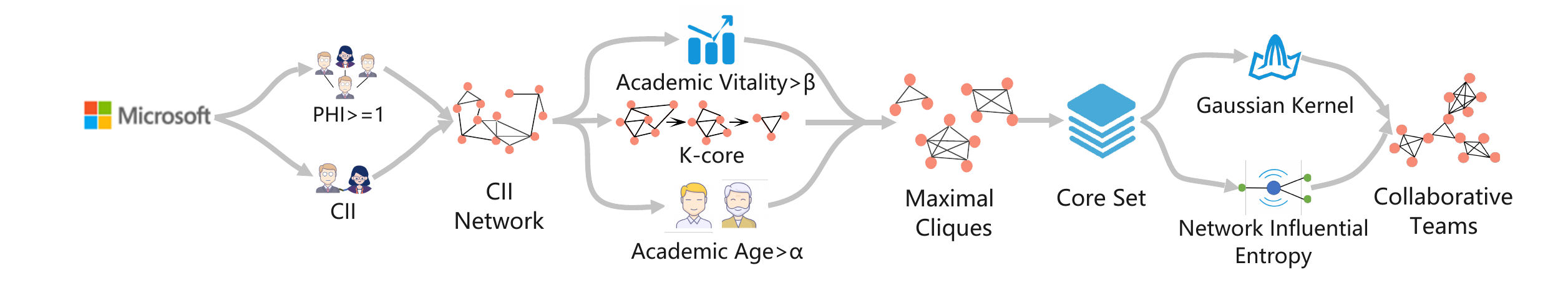}
	\caption{CORE framework. With the input co-author network, CORE consists of three steps: (1) constructing CII network, (2) finding core teams, and (3) recognizing collaborative teams. Different labels are then assigned to nodes in different collaborative teams. Subgraphs with the same labels and edges between them are considered recognized teams.}
	\label{alstructure} 
\end{figure}
	
	The basic concept of the CORE algorithm can be summarized in three steps, as shown in Figure \ref{alstructure}. First, we construct the CII network, whose edges are weighted by the CII. Moreover, scholars with sole academic behaviors and inactive collaborative behaviors were excluded. Then, we find the core team members according to their structural significance and academic behaviors. We employed the \textit{k-core decomposition} method to capture nodes with high degree values and measured their academic behaviors according to two metrics: academic age and academic vitality. Based on the recognized core team members, we optimized the network influential entropy, which reflects the teams' structures in their local areas, to find the entire collaborative teams. After these three steps, we can identify collaborative teams with high-quality cores and extensions whose members collaborate closely with core members within collaborative teams.

\subsection{Constructing CII Network}
	Algorithm \ref{alg:1} presents the pseudocode of the first step of CORE in constructing the CII network. The input of this step is the co-author network, and the output is the new CII network. Specifically, $ G=(V_w,E_w) $ is the original co-author network with weighted node set $ V_w $ whose node weights are the numbers of publications, and the weighted edge set whose edge weights are the numbers of co-author publications. $ G_{CII}=(Vc_{w},Ec_{w}) $ is a network with a weighted edge set whose edge weights are CII values between pairs of nodes. We first calculated the PHI for each scholar based on their neighbors in the network. According to the definition of PHI, scholars with $ \varphi > 1 $ can be regarded as having active collaborative relationships. By contrast, $ \varphi = 1 $ for a scholar means that they have collaborations with only one scholar or only one collaboration with any number of other scholars. Scholars with $ \varphi = 0 $ are the isolated nodes in the co-author network. We consider co-author behaviors as collaborative relationships so that isolated nodes that have no neighbors are considered as scholars who never collaborate with other scholars. Those whose $ \varphi = 1 $ have simplex collaboration relationships (only collaborate with one scholar) or sparse collaboration relationships (collaborate with many scholars once). Thus, we removed these nodes because they barely contributed to the collaborative team recognition.

	\begin{algorithm}[ht]
		\caption{Constructing CII Network}
		\label{alg:1}
		\begin{algorithmic}[1]
			\REQUIRE ~~\\
			$ G=(V_w,E_w) $
			\ENSURE ~~\\
			$ G_{CII}=(Vc_{w},Ec_{w}) $
			\FOR{each $ V \in V_w $}
			\STATE Obtain edge set $ s(V) $ with edges containing node $ V $
			\STATE Calculate $ PHI(V) $ according to $ s(V) $
			\IF{$PHI(V) \leq 1$}
			\STATE Remove $ V $ from $ V_w $
			\ELSE
			\STATE Obtain node $ V $'s neighbor node set $ n(V) $
			\FOR {$ N $ in $ n(V) $}
			\STATE $ E_{c_w}(N, V) = \Delta k _{NV}^{2} / (\Delta k_N \Delta k_V)$
			\ENDFOR
			\ENDIF
			\ENDFOR
			\RETURN $ G_{CII} $
		\end{algorithmic}
	\end{algorithm}
	
	Then, we calculated the CII between all pairs of nodes in the network as edge weights to evaluate the collaborative relationships. Finally, we removed the isolated nodes and edges with no neighbors to construct the CII network.
	
\subsection{Finding Core Teams}
	In this step, we used two parameters (i.e., $\alpha$ and $\beta$) to restrict scholars with active collaborations during the time window to recognize collaborative teams. We adopt \textit{k-core decomposition}, which is a coarsening method for ranking the importance of nodes. By successively removing nodes whose degrees are less than \ textit{k}, we can find the node set in which the node degrees are greater than or equal to \textit{k}. However, the teams should be composed of more than two people. We apply the function $ k\_core( G_{CII},k=3) $ to find the 3-core subnets $ k\_sub $ in $ G_{CII} $ as core candidates. Then, we remove scholars whose academic age is less than $\alpha$ and whose academic vitality is less than $\beta$ from $ k\_sub $ to obtain the final core members. To capture densely connected subgraphs, including core members, we find the maximal cliques for all core members. The maximal clique for node \textit{v} is defined as \textit{MC(v)}, which is the largest complete subgraph containing the given node \textit{v}. We calculated the maximal cliques for all eligible nodes in $ k\_sub $ and then added their node sets to $ \phi $.
	
	We noticed that some core nodes have more than one maximal clique. Therefore, some node sets in $ \phi $ are highly similar to others. We measured the similarity between two node sets using their Jaccard similarity coefficient (JSC), which is calculated as follows:
	\begin{equation}
		J(A,B)=\frac{|A \cap B|}{|A \cup B |},
		\label{jac}
	\end{equation}
	where $ A,B $ are two different sample sets. JSC is a commonly used metric to measure overlaps in two sets, which increases if the two sets have more common elements. In this step, we merged two maximal cliques into one team if the JSC between them was larger than $ \varepsilon $. We empirically set $ \varepsilon = 0.5 $, that is, two maximal cliques of a node \textit{v} own the same half of all nodes.

	We calculate the JSC between maximal cliques in $ \phi $ and union of two maximal cliques together if the JSC between them is greater than 0.5. The aforementioned processes aim to achieve node sets as core teams. After these procedures, we find the core teams $ \Omega  = \{ Core_1,Core_2,...,Core_n \}$ of all the collaborative teams. The details of finding the core teams are given in Algorithm \ref{alg:2}.
	\begin{algorithm}[ht]
		\caption{Finding Core Teams}
		\label{alg:2}
		\begin{algorithmic}[1]
			\REQUIRE ~~\\
			$ G_{CII}=(Vc_{w},Ec_{w}) $
			\ENSURE ~~\\
			$ \Omega  = \{ Core_1,Core_2,...,Core_n \}$
			
			\STATE $ k_sub \gets k\_core( G_{CII},k=3) $
			\STATE $ \Omega  \gets \varnothing$
			\STATE $ \phi \gets \varnothing$
			\FOR {$ V \in  k_sub $}
			\STATE $ Aa(V) \gets $ Academic Age of $ V $
			\STATE $ Av(V) \gets $ Academic Vitality of $ V $
			\IF {$ Aa(V) > \alpha$ and $ Av(V)> \beta$}
			\STATE $ MC(V) \gets $ Maximal Clique of $ V $
			\STATE Add $MC(V)$ to $ \phi $
			\ENDIF
			\ENDFOR
			\FOR {$ (MC_i, MC_j)\in \phi $}
			\STATE Calculate $ J(MC_i, MC_j)$ by Eq.~\eqref{jac}
			\IF {$ J(MC_i, MC_j) > \varepsilon $}
			\STATE $ Core \gets MC_i \cup MC_j $
			\STATE Add $ Core $ to $ \Omega $
			\STATE Remove $ MC_i, MC_j$ from $\phi$
			\ENDIF
			\ENDFOR
			\STATE $\Omega \gets \phi \cup \Omega$
			\RETURN $\Omega $
			
		\end{algorithmic}
	\end{algorithm}	
\subsection{Recognizing Collaborative Teams}
\begin{figure}
		\centering
		\includegraphics[width=0.6\textwidth]{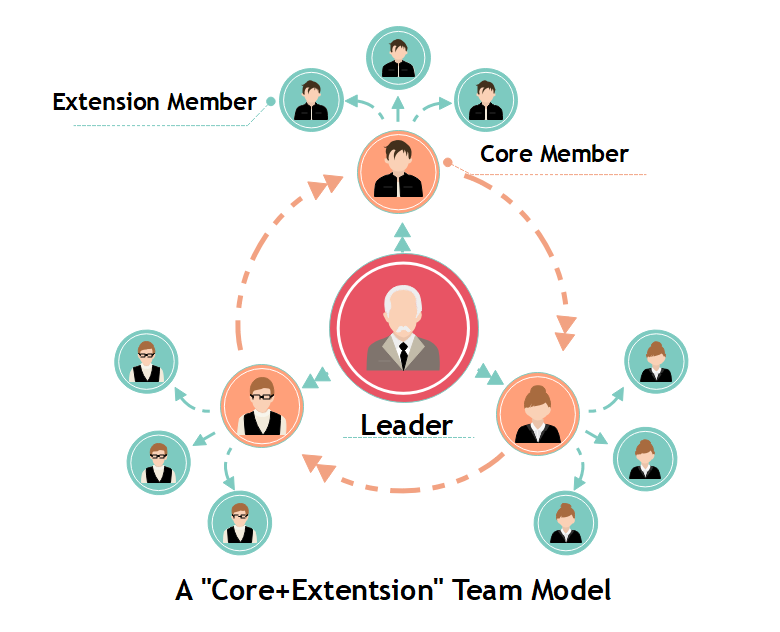}\\
		\caption{"Core + extension" team structure.}
		\label{fig:c+e}
	\end{figure}

After identifying the core teams in the network, we determine the entire collaborative teams by finding extensions that are closely connected to the cores. Owing to the structure of the collaborative teams shown in Figure~\ref{fig:c+e}, we add nodes with the following properties to the core teams:
	\begin{enumerate}
		\item Core members are well connected to their extension members. However, it is not necessary that extension members are in close relationships with each other.
		\item Core members generally have higher influence in the team compared with extension members.
		\item Core members spread collaboration influence on their own relevant followers. Combining with clique structures among core members, cores, and their followers finally construct stable inner structure for their teams.
	\end{enumerate}

	Based on the properties mentioned above, we employed the Gaussian kernel function and network influential entropy to recognize all collaborative teams. Gaussian kernel function is a commonly used kernel function, which can map finite dimensional data to high-dimensional space so that the originally linearly inseparable data can be linearly separable. The Gaussian kernel function primarily involves the calculation of the Euclidean distance between two vectors, and changing $\sigma$ can control the local action range of the kernel function. First, we change the distance function to CII such that the Gaussian kernel function ${K(i)}$ can measure the cooperative leadership of core members. Thus, ${K(i)}$ not only obtains the influence of cooperation but also ensures connectivity between core and general members. In the previous section, we used the maximum complete subgraph to ensure that team members were fully identified. This solved the problem in which the k-core decomposition algorithm could not guarantee that the maximum k-kernel network is the densest subnetwork in a given network to a certain extent. The collaborative influence of \textit{i} is calculated based on all their collaborators. Thus, the collaboration influence distribution function of member \textit{i} is defined as
	\begin{equation}
		K(i)=\sum_{j\in n(i)} \exp\! \frac{-(1-CII_{ij})^2}{2\sigma^{2}},
		\label{KI}
	\end{equation}
	where $ CII_{ij} $ is the CII between \textit{i} and \textit{j}, $ \sigma \in (0,+ \infty) $ is the influence factor controlling the interactive field of core member \textit{i}, and $ n(i) $ is the set of collaborators of \textit{i}, which is the neighbor node set of \textit{i} in the CII network. A larger $ \sigma $ means that \textit{i} has a broader interactive field, which indicates that \textit{i} maintains close collaborative relationships with more of their collaborators.
	
	The PauTa criterion demonstrates that the Gaussian function $ f(x) $ rapidly reduces to 0 when $ x $ is larger than $ \mu+3\sigma $. Thus, the interactive field of each node is approximately $ 3\sigma $ controlled by influence factor $ \sigma $. When $ (1-CII_{ij}) \leq 3\sigma $, the exponential function $ K(i) $ is greater than 0. By contrast, when $ (1-CII_{ij}) $ is greater than $3\sigma $, the exponential function $ K(i) $ rapidly decreases to 0. The scholars in $ n(i) $ whose $ CII \geqslant 1- 3\sigma $ construct the interactive field of core members \textit{i}, and those whose $ CII < 1- 3\sigma $ with core member \textit{i} are fringe collaborators, who can be identified as nodding acquaintances of \textit{i}. As a result, we can find scholars in the interactive field of core members who collaborate closely with cores. Moreover, core members have the highest influence on their own interactive fields.
	\begin{algorithm}[ht]
		\caption{Recognizing Collaborative Teams}
		\label{alg:3}
		\begin{algorithmic}[1]
			\REQUIRE ~~\\
			$ G_{CII}=(Vc_{w},Ec_{w}) $\\
			Core Communities $ \Omega = \{ Core_1,Core_2,...,Core_n \}$
			\ENSURE ~~\\
			Resulting clusters $ \psi = \{C_1, ..., C_n\} $
			
			\STATE $ \psi  \gets \varnothing$
			\FOR {$ Core \in \Omega $}
			\STATE $ C_{team} \gets \varnothing $
			\STATE $ V_{team} \gets Core$
			\FOR {$ p \in Core $}
			\STATE Obtain $ G_{p}= (V_{p},E_{p}) $ the ego network of $ p $
			\STATE Calculate $ f(p,\sigma) $
			\STATE Obtain $\sigma$ by solving Eq. \eqref{Fmodel}
			\STATE $ CII_{\sigma} \gets 1- 3 \sigma $
			\FOR {$ edge \in E_{p} $}
			\IF {$ Weight(edge) < CII_{\sigma} $}
			\STATE Remove $edge$ from $ G_{p} $
			\ENDIF
			\STATE Remove isolated nodes from $ G_{p} $
			\ENDFOR
			\STATE $ V_{team} \gets V_{team} \cup V_p $
			\ENDFOR
			\STATE $ C_{team} \gets Subgraph(V_{team}) $
			\STATE Add $ C_{team} $ to $\psi$
			\ENDFOR
			\RETURN $\psi$
		\end{algorithmic}
	\end{algorithm}
	Clearly, the influence factor $ \sigma $ controls the interactive field of a node, which determines the team recognition results. That is, we can use $\sigma $ to control the team size. Here, we define the network influential entropy to adjust the proper values of $ \sigma $ for the different core nodes in the network. Network influential entropy has been proposed to measure network stability by combining network topology with information theory~\citep{cai2017analysis}. We propose network influential entropy to quantify the capability of cores to achieve stable collaborations in terms of collaboration leadership. Let $ G_{i} =(V_{G_{i}},E_{G_{i}}) $ be the ego network of node \textit{i}, and define the network influential entropy $ H_{i} $ in Eq. \eqref{HI}, which takes nodes and edges in the ego network into account.
	\begin{equation}
		H_i=-\sum_{v{\in}V_{G_{i}}}\frac{K(v)}{\sum_{v{\in}V_{G_{i}}}K(v)} \log \frac{K(v)}{\sum_{v{\in}V_{G_{i}}}K(v)}
		\label{HI}
	\end{equation}
	
	In Eq. (\ref{HI}), $ V_{G_{i}}$ is the node set of the ego network of $ i $, and $ K(v) $ is the collaborative leadership distribution function of node $ v $ which is defined in Eq. \eqref{KI}. Note that a smaller $ H_i $ value represents a more stable collaboration network of node \textit{i}. Thus, we can minimize the network influential entropy $ H_i $ to determine the most stable collaboration network of node \textit{i}. When we fix $ i $ in the ego network of $i$ and consider $\sigma $ as a variable, Eq. \eqref{KI} can be changed as follows:
	\begin{equation}
		f(i,\sigma) = \sum_{j=1}^{n} \exp\left(\frac{c_j}{2\sigma^2}\right),
		\label{simple_f}
	\end{equation}
	where $ n $ is the size of the neighbors of $ i $, and $ c_j $ is constantly calculated by $-(1-CII_{ij})^2$. We substitute Eq. \eqref{simple_f} to Eq. \eqref{HI}, turning the problem into the following optimization model:
	\begin{equation}
		\begin{split}
			& \underset{\sigma}{\min}-\sum_{v \in V_{G_{i}}}\frac{f(v,\sigma)}{\sum_{v \in V_{G_{i}}}f(v,\sigma)} \log \frac{f(v,\sigma)}{\sum_{v \in V_{G_{i}}}f(v,\sigma)}\\
			& \rm s.t. \quad 0 < \sigma \leq \frac{1}{3}.
		\end{split}
		\label{Fmodel}
	\end{equation}
	
	There are several existing methods for optimizing the above single-objective function, such as simulated annealing, genetic algorithms, and ant colony algorithms. Specifically, if $ \sigma $ is equal to $\frac{1}{3} $, the nodes interact with all collaborators. We calculate $\sigma$ for all members in the same core team to obtain their extensions and then unite them to find complete teams for the given core team.
	
	We input the resulting CII network $ G_{CII} $ from Algorithm \ref{alg:1} and the core teams $ \Omega $ calculated from Algorithm \ref{alg:2}. For a given core community in $ \Omega $, we obtain the ego networks of each node inside the core community and then calculate the proper $ \sigma $ by solving Eq. \eqref{Fmodel}. We calculate $ CII_{\sigma} $ according to the obtained $ \sigma $, remove edges in $ G_{p} $ (the ego network of $ p $) whose weights are less than $ CII_{\sigma}  $, and then update its node list $ V_{p} $ by removing the isolated nodes. Finally, we create a union of all node lists in the core team as $ V_{team} $ and let the subgraph $ C_{team} $ containing $ V_{team} $ be a resulting cluster. We repeat these procedures for all core teams in $ \Omega $ to obtain the resulting clusters $ \psi $ in the network. Details of this step are presented in Algorithm \ref{alg:3}.

\subsection{Complexity Analysis}
The time complexity of each step is summarized in Table \ref{Complex}. The construction step requires computing all node and edge weights in the graph, which takes $ O( |V_{w}|+ |E_{w}| ) $. According to the definition of the k-core and clique of the network, we should take $ O(|(Vc_{w}|^2) $ to obtain core candidates as the max clique set $ \phi $ and then take $ O(2^{|\phi|}) $ to find core communities in the network. Finally, we optimized Eq. \eqref{Fmodel} $|\Omega|$ times, which takes $ O(|Vc_{w}| \times |\Omega|) $. From the experiment, we found that $ |\Omega| $ is approximately equal to $ |\phi| $, which is much smaller than $|Vc_{w}|$($|\phi|\approx \frac{1}{20}|Vc_{w}|$). Thus, the complexity of our method can be simplified to $O(|V_{w}|+ |E_{w}|+|Vc_{w}|^2+2^{|\phi|})$.
		\begin{table*}[ht]
		\caption{Time complexity for each step.}
		\begin{tabular}{ccc}
			\hline
			Step                            & Sub-step                                                    & Computational Complexity \\ \hline
			Constructing CII \\Network        & ---                                                         & $ O(|V_{w}|+ |E_{w}|)     $                   \\ \hline
			Finding Core \\Teams        & \begin{tabular}[c]{@{}c@{}}Filtering\\ Merging\end{tabular} & \begin{tabular}[c]{@{}c@{}} $ O(|(Vc_{w}|^2)$\\  $ O(2^{|\phi|}) $\end{tabular}\\                        \hline
			Recognizing \\Collaborative Teams & ---                                                         &$ O(|Vc_{w}| \times |\Omega|) $\\\hline
		\end{tabular}
		\label{Complex}
	\end{table*}

\section{Experimental Results}~\label{sec4}
In this section, we elaborate on some details of our experiments, that is, dataset information, data preprocessing, experimental environment, and parameter settings. Then, we evaluate the performance of the CORE method by comparing it with state-of-the-art methods. We adopted collaborative team evaluation methods as performance metrics to evaluate team recognition performance. Moreover, we explored collaboration patterns from a scientific perspective based on the resulting teams.
\subsection{Datasets}
\begin{figure}
  \centering
  \includegraphics[width=0.6\textwidth]{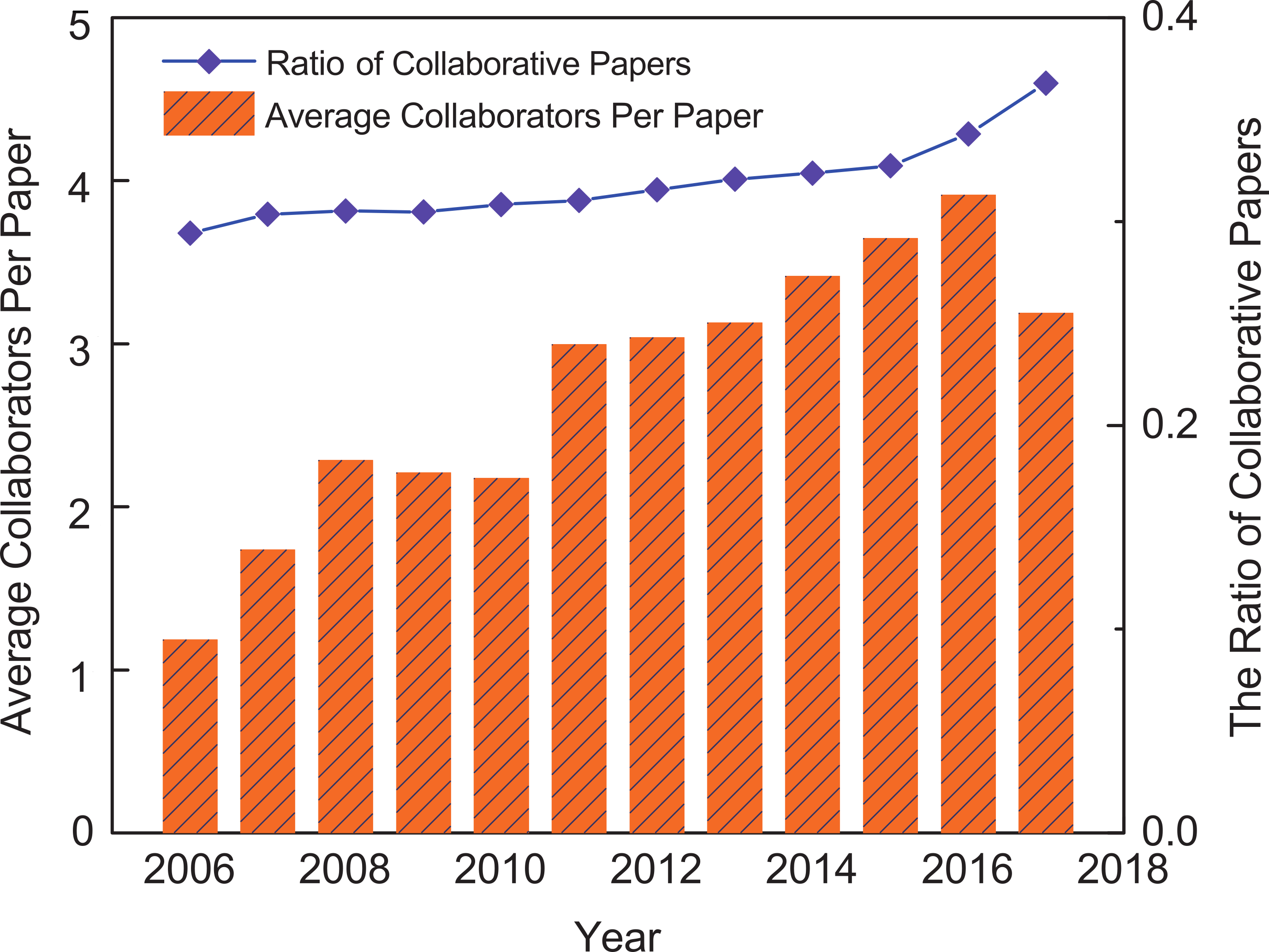}
  \caption{Average collaborators per paper and the collaborative publication ratio in MAG-CS from year 2006 to 2017.}
  \label{fig:acpc}
\end{figure}

\begin{figure}
  \centering
  \includegraphics[width=0.6\textwidth]{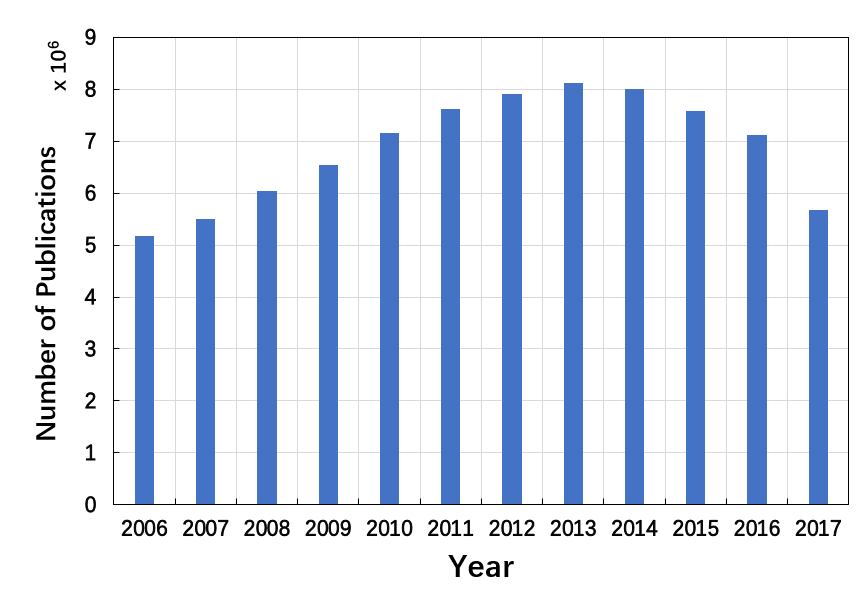}
  \caption{Distribution of publications over time.}
  \label{fig:pubdis}
\end{figure}

To recognize teams and reveal collaboration patterns, we use the scholarly dataset Microsoft Academic Graph (MAG)\footnote{\url{https://www.microsoft.com/en-us/research/project/microsoft-academic-graph/}} provided by Microsoft Academic Service~\citep{2015MAS}, which contains various information about scholars and their scientific outcomes. It is made up of more than 50 million authors and 90 million publications from 1800 to 2017. The MAG-CS data sets contain research publications with 19 main research fields, computer science, physics and biology, etc. In this study, we primarily focus on the computer science publication dataset, which contains a total of 9,987,172 publications and 12,552,004 scholars. Figure~\ref{fig:acpc} shows the basic statistical information of collaborations in the selected publications. Figure~\ref{fig:pubdis} shows the distribution of the publications over time.
%


A long time window helps scholars to find stable collaborations. At the same time, it also makes it more reasonable for collaborative teams recognition. We analyzed the collaboration durations between every pair of scholars. It was found that more than 96.02\% of scholars will not collaborate again if they have not collaborated for 4 years since their last collaboration. Among them, some collaborators may never publish papers after a 4-year interval, while others may continue their academic careers. Therefore, we used 4 years as a time window and 2 years as a movement to identify collaborative teams.

We considered the nearest 12 years as a phase to generate co-author networks in each time window, that is, 2006--2009, 2008--2011, 2010--2013, 2012--2015, and 2014--2017. In addition, we considered scholars' academic careers and educational years in reality and set $ \alpha = 5 $ ~\citep{wang2017scientific, yu2017team}. Based on real-world observations, scholars are often in their student stages if their academic age is less than 5, often working with their supervisors. We chose $ \beta = 1 $, that is, scholars publish one paper every year, to guarantee that core scholars play an active role in their teams.

\subsection{Comparison Methods}
In this work, we chose 6 algorithms as the baseline methods: CPM, Linkcomm, LFM, SLPA, BigClam, and Demon. These algorithms are elaborated on below.

\begin{itemize}
  \item \textbf{CPM}~\citep{palla2005uncovering} --- A classical clique percolation algorithm, which hypothesizes that a community is composed of overlapping sets of complete subgraphs and detects communities by searching adjacent cliques. This algorithm constructs a new graph of all cliques of size \textit{k}, whose connected components determine how cliques compose the communities.

  \item \textbf{Linkcomm}~\citep{ahn2010link} --- A link partitioning algorithm, which discovers community structures by partitioning links in the network. The Linkcomm algorithm partitions links based on their edge similarities and then uses hierarchical clustering method to yield links for different communities.

  \item \textbf{LFM}~\citep{lancichinetti2009detecting} --- A local expansion algorithm, which utilizes a local benefit function to grow a community from a group of high-quality nodes. The LFM algorithm yields a community from a random seed node $ c $ by optimizing the defined fitness function $ f(c) $ until $ f(c) $ reaches the local maximum.

  \item \textbf{SLPA}~\citep{xie2011slpa} --- An agent-based dynamic algorithm, which assigns different labels for nodes and gathers nodes with same labels together as a community. The SLPA algorithm allocates speaker-listener labels for different nodes through information propagation processes, which spread labels between nodes based on both previous and current iterative information of nodes.

  \item \textbf{BigClam}~\citep{yang2013overlapping}--- A fuzzy detection algorithm, which calculates belonging or membership vectors for every node and then uses matrix factorization-based methods to partition communities. The BigClam algorithm builds a bipartite affiliation network to model community structures. According to the new bipartite graph, it utilizes underlying adjacency matrix of the given graph to maximize the likelihood of a new nonnegative matrix $ F $ which represents the probabilities for each edge.

  \item \textbf{Demon}~\citep{coscia2012demon}--- A seed expansion algorithm, which expands random seeds to subsets and then identifies communities based on these subsets using other algorithms. The Demon algorithm expands each seed to a subspace and adopts the label propagation algorithm ~\citep{raghavan2007near} to identify nonoverlapping communities. Then, the Demon formulates final overlapping communities by combining these nonoverlapping communities in different subspaces.
\end{itemize}

In our experiments, we set $ k=3 $ for the CPM algorithm and $ \alpha= 0.8 $ for the LFM algorithm. In the SLPA algorithm, we used the maximum iterations $  T = 10 $ and label probability $ r = 0.1 $. For the BigClam algorithm, we set the community number to be $ n_c $, which is the number of collaborative teams in our algorithm. In the Demon algorithm, we set $ \epsilon = 0.25 $, and the minimal community size was 3. All comparison experiments were conducted on a desktop with an Intel Core i5-6500 CPU, 16G RAM, and Windows 10 operating system.

\subsection{Evaluation Metrics}
In addition to evaluating collaborative teams in terms of scientific outcomes, such as team papers, citations, and diversity impacts, we investigated teams according to team structure. As mentioned above, collaborative teams are a type of overlapping cluster in a graph. Clustering validation measures can be adopted to understand internal structures and external connections with their neighboring clusters. Due to the lack of real team datasets, we evaluated teams using scoring functions instead of comparing them with the ground truth. Nearly all commonly used scoring methods are based on the intuition that a good cluster has dense connections with internal nodes and sparse connections with the remaining nodes. Similarly, we can mathematically formalize this intuition to quantify the desirable properties of collaborative teams. Considering the structure of collaborative teams, we adopted five types of goodness metrics, which focus on various aspects of graph structures, to evaluate collaborative teams~\citep{paudel2020approach}.

\begin{itemize}
  \item \textbf{Triangle motif} is a basic phenomenon in social circles, which reflects the probability of two acquaintances becoming friends. In a  collaboration network, we use triangles in collaborative teams to check ternary closures, meaning that we calculate the number of triangles for nodes to characterize internal structure. If a collaborative team has more triangles, it generally means that members have dense connections with each other.
  %
  \item \textbf{Average path (AvgPath)} is a direct metric to measure how tight the inner structures of collaborative teams are. AvgPath of the team \textit{c} with connected graph structure is defined as the average value of the shortest path length between pairs of nodes in a team, which is calculated as
  \begin{equation}
    AvgPath(C) = \frac{2 \times \sum_{u,v}^{c} Shortest Path(u,v)}{|c|(|c|-1)},
  \end{equation}
  where $ Shortest Path(u,v) $ is the shortest path length between $ u $ and $ v $, and $ |c| $ is the number of team members in team $ c $. The shortest path length can reflect closeness between two nodes; therefore, members in teams with high AvgPath maintain close relationships with each other. In other words, such teams have dense internal connections.

  \item \textbf{Separability (Sep)} satisfies the intuition that good teams are well-separated from the rest of authors in a co-author network. Formalizing the intuition from network connection views, a good collaborative team \textit{C} has few connections pointing from members in the team \textit{C} to the remaining authors in a co-author network. Separability is defined as
  \begin{equation}
    Sep(C)=\frac{Out_{C}}{All_{C}},
    \label{sep}
  \end{equation}
  where $ Out_{C} $ is the number of edges on the boundary of \textit{C}, and $ All_{C} $ is the number of edges connected by nodes in \textit{C}. Sep measures the ratio between external number of connections and the number of all connections by team members.

  \item \textbf{Clustering coefficient (CCF)} builds on the intuition that good teams have compact internal structure, with members intending to collaborate closely with each other. Reflecting on the distribution of connections in co-author networks, pairs of nodes with more common neighbors are more likely to connect with each other. In other words, the CCF of a node is the fraction of possible triangle motifs through that node. CCF for team \textit{C} is the average value for all members' clustering coefficients inside the team, which is defined as
  \begin{equation}
    CCF(C)=\frac{1}{\left |C  \right |}\sum_{u\epsilon C}\frac{2T(u)}{deg(u)(deg(u)-1)},
  \end{equation}
  where $ |C| $ is the number of team members, \textit{T(u)} is the number of triangles through node \textit{u}, and \textit{deg(u)} is the node degree of \textit{u}.

  \item \textbf{Internal community degree fraction (ICDF)}~\citep{yang2015defining} is a metric satisfying not only the intuition that well-functioning teams have dense connections but also the structure that teams are assembled by cores with extensions. Intuitively, a  well-functioning team is expected to have high internal conductance. Cores play an important role as connectors in a team, which promotes internal conductance of the team. We employed the ICDF of node \textit{u} in the collaborative team \textit{S} to evaluate internal conductance. The ICDF is defined as
  \begin{equation}
    ICDF(u,S)=\frac{\left | \left \{ v|v \epsilon S, (u,v)\epsilon E_{S} \right \} \right |}{|S|},
  \end{equation}
  where $ E_{S} $ are the edges in team \textit{S}, and $ |S| $ is the number of team members. The scoring function for a collaborative team \textit{C} is the maximum value of ICDF for nodes in the team, which is defined as
  \begin{equation}
    g(C)= \max_{u \epsilon  s}ICDF(u,C).
  \end{equation}
  A team with a larger value of $g(C)$ shows that there exists a member who connects more team members. Due to the existence of this member, a collaborative team will have a denser internal structure. Moreover, existing members with maximum values of ICDF often appear as core members in teams.
\end{itemize}
\begin{figure*}[!htbp]
  \centering
  \subfigure[2006-2009]{
    \includegraphics[width=0.30\textwidth]{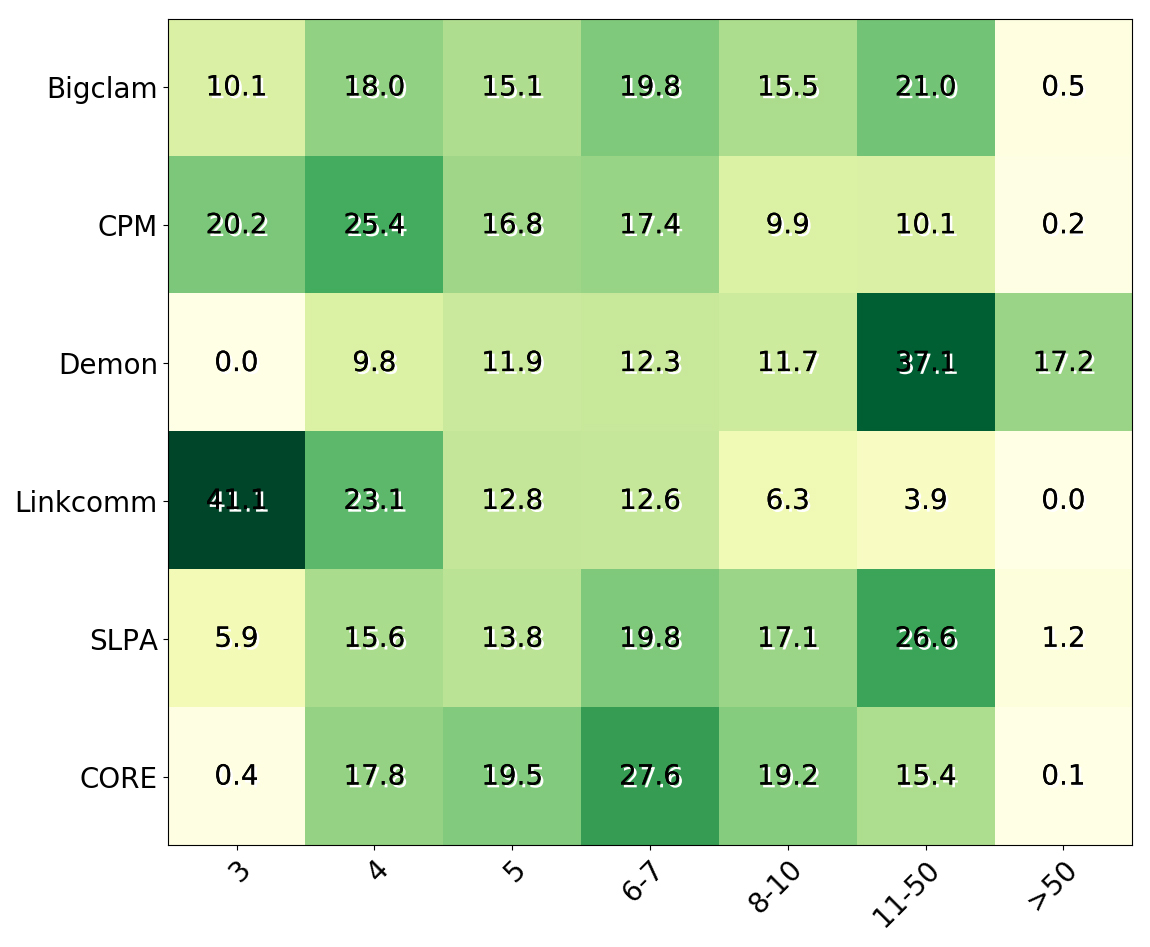}
  }
  \subfigure[2008-2011]{
    \includegraphics[width=0.30\textwidth]{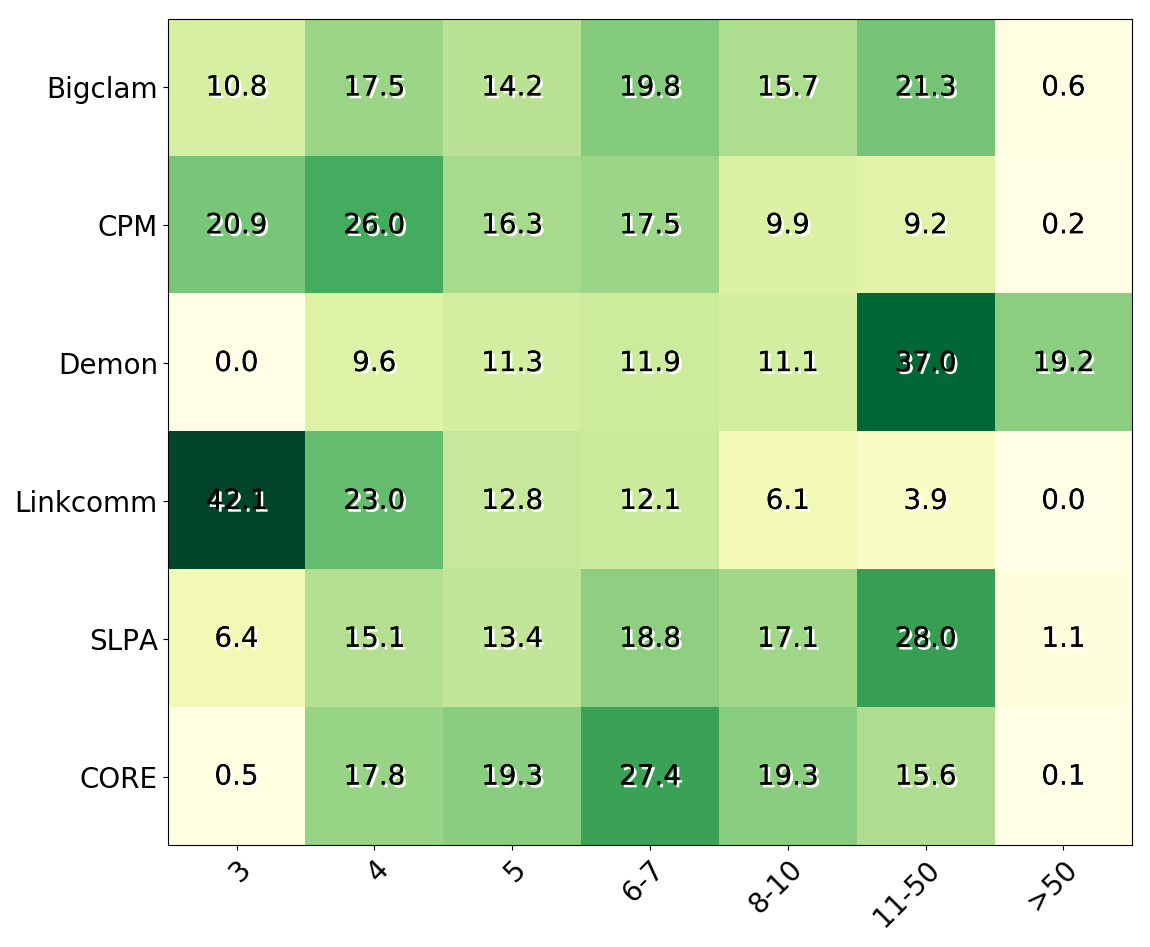}
  }
  \subfigure[2010-2013]{
    \includegraphics[width=0.30\textwidth]{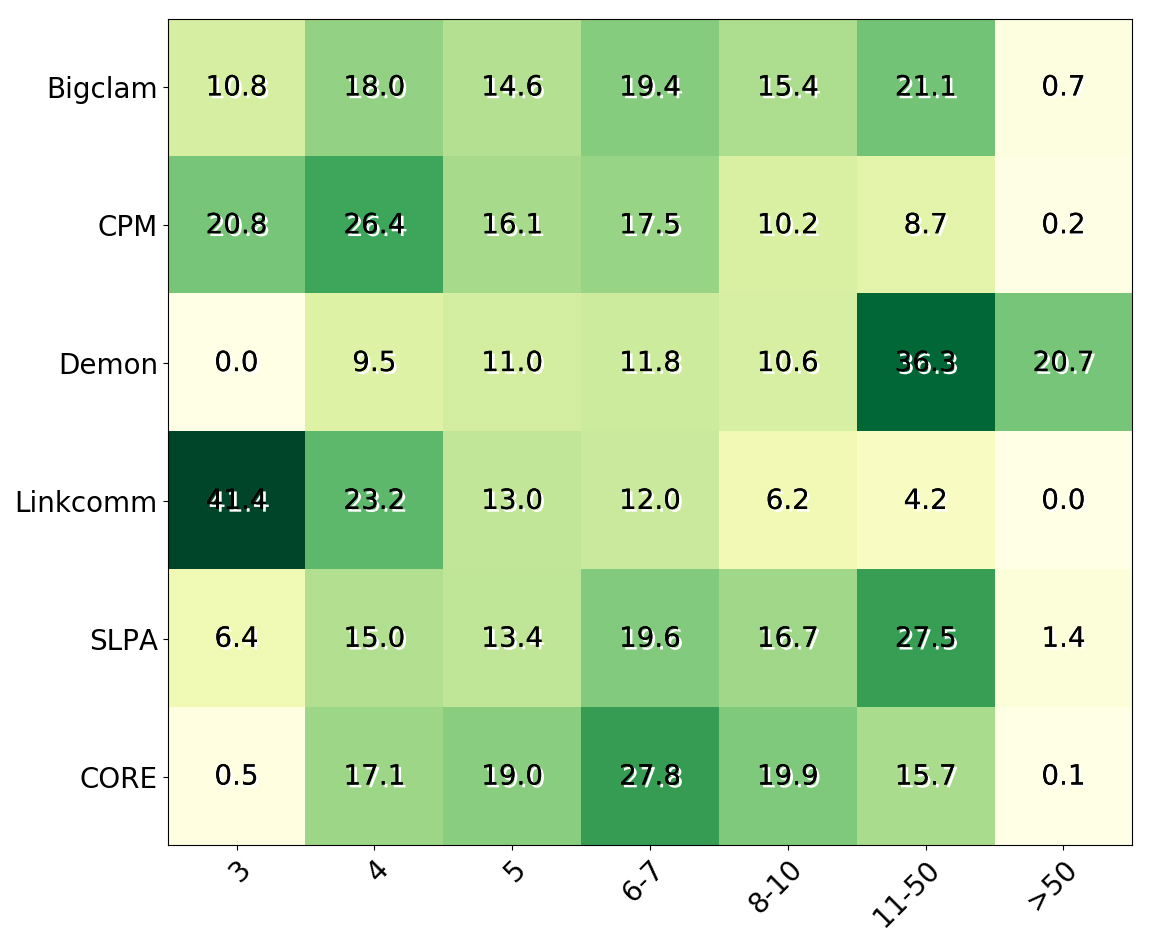}
  }\\
  \subfigure[2012-2015]{
    \includegraphics[width=0.30\textwidth]{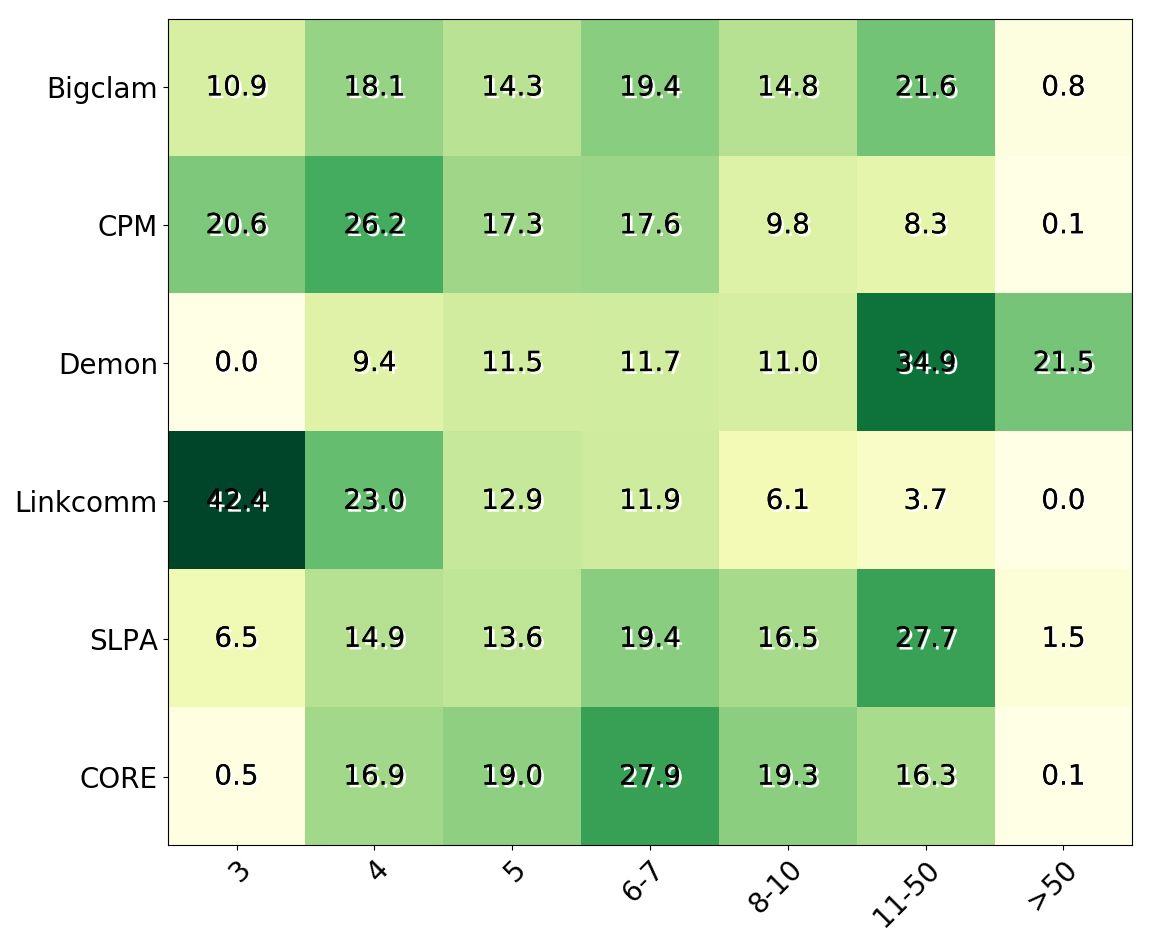}
  }
  \subfigure[2014-2017]{
    \includegraphics[width=0.30\textwidth]{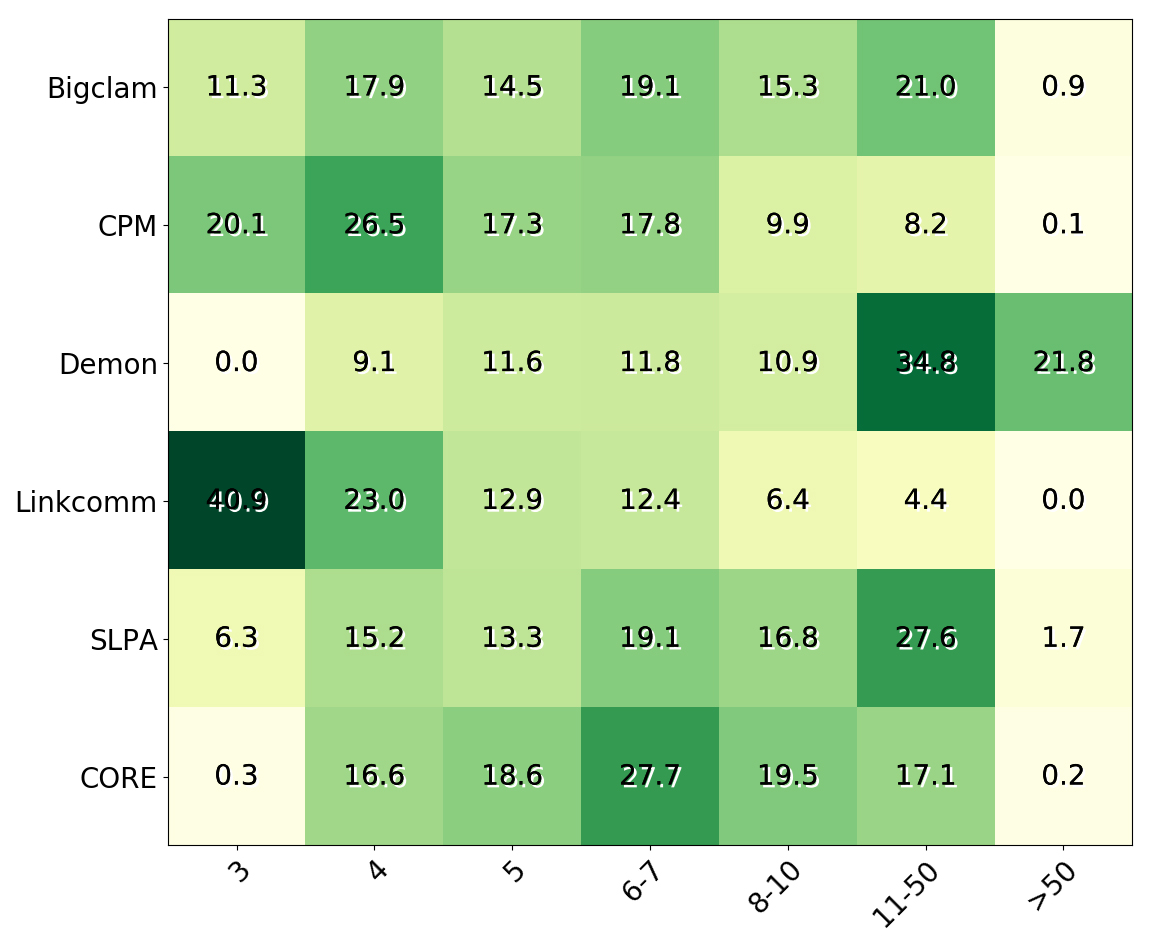}
  }
  \subfigure[Number of recognized teams]{
    \includegraphics[width=0.33\textwidth]{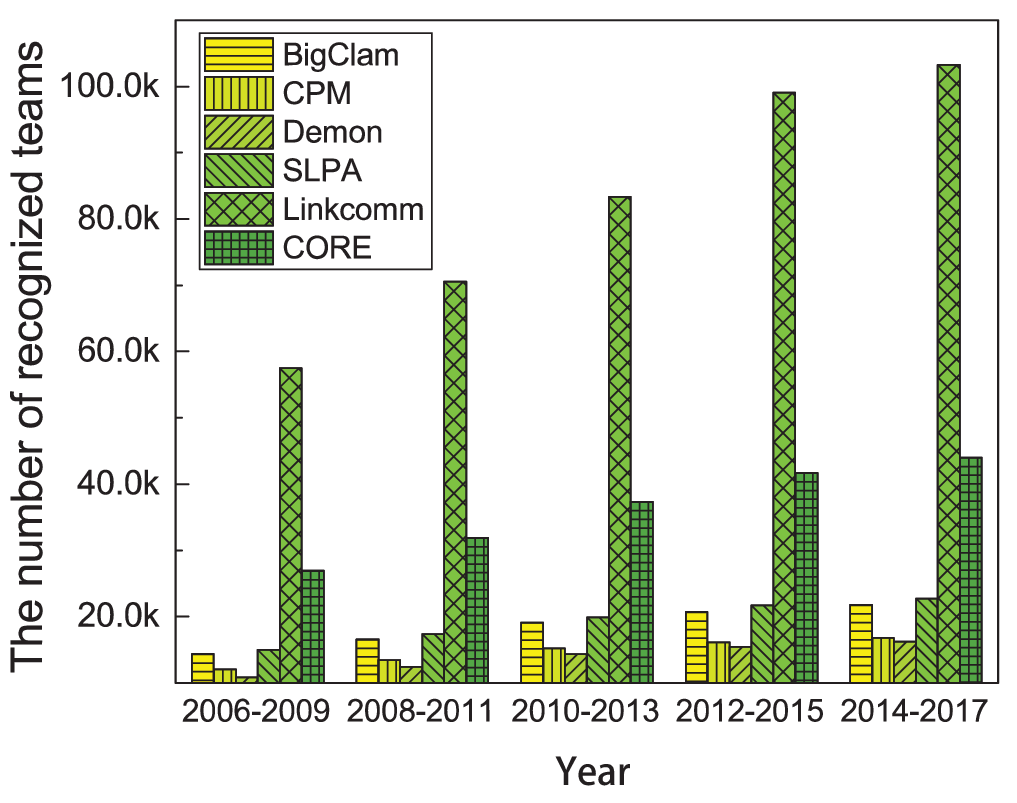}
  }
  \caption{Team-scale distributions identified by CORE and other baseline algorithms. In (a) to (e), the numbers represent the percentages of recognized team scales for each method, and darker colors refer to higher proportions. In (f), the general distributions of recognized teams are shown.}
  \label{fig:memdis}
\end{figure*}
\subsection{Evaluation of CORE}
	
Figure \ref{fig:memdis}(a)--\ref{fig:memdis}(e) show the team-scale distributions for the different methods in different time windows. The columns represent the scales of the identified teams and the rows represent the different algorithms. Each cell contains percentages of the different team scales. For Linkcomm, 41.1\% of the recognized teams have 3 members. For Demon, there are no teams with 3 members, whereas over 50\% of the recognized teams have more than 11 members. This is also observed in Figure~\ref{fig:memdis}(f). Demon recognized the lowest number of teams (10,862 in 2006--2009, 12,430 in 2008--2011, 14,355 in 2010--2013, 15,421 in 2012--2015, and 16,244 in 2014-2017), whereas Linkcomm recognized the most (57,503 in 2006--2009, 70,578 in 2008--2011, 83,323 in 2010--2013, 99,901 in 2012--2015, and 103,242 in 2014--2017). This indicates that the recognized teams of Demon are too large, whereas those of Linkcomm are too small.

As for other baseline methods, such as BigClam, SLPA, and CPM, the recognized collaborative teams are relatively regular in size. According to statistics~\citep{fortunato2018science}, in computer science research, the average team size is approximately 6. Compared with baseline methods, CORE can yield better overall results for team recognition. As shown in Figure~\ref{fig:memdis}(a) to (e), most teams are of the size of [6, 7] in CORE. It is obvious that the scales of collaborative teams recognized by Demon are generally too large while those of Linkcomm are generally too small. With the passage of time, the sizes of all teams gradually increase, which is reflected in Figure~\ref{fig:memdis}(f). We can see that our proposed CORE algorithm can achieve reasonable and accurate results, which agree with existing sociological theories and team science developments~\citep{fortunato2018science}.

\begin{figure*}[!tbp]

  \centering
  \subfigure[Average team size]{
    \includegraphics[width=0.30\textwidth]{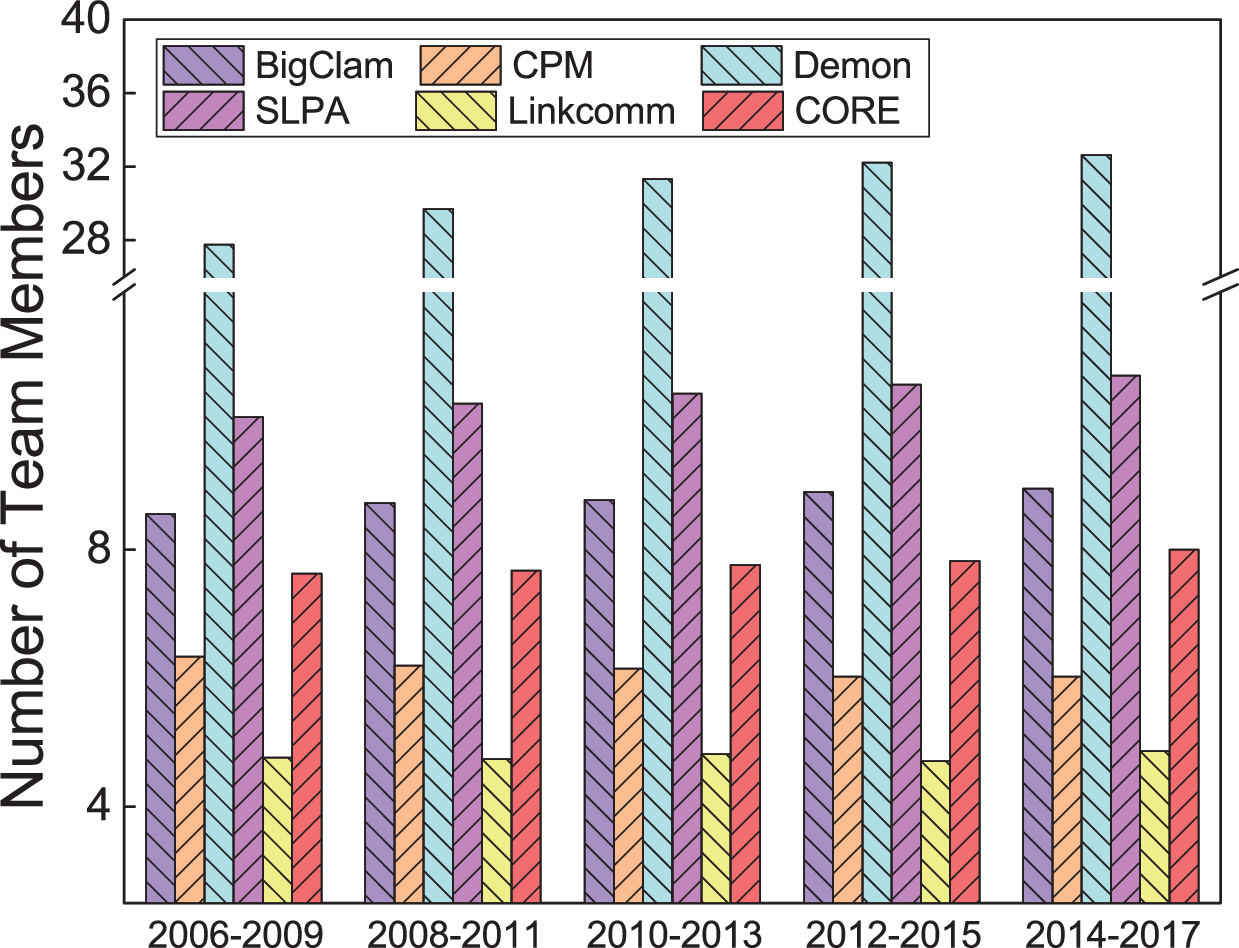}
    \label{fig:avgnum}
  }
  \subfigure[Separability]{
    \includegraphics[width=0.30\textwidth]{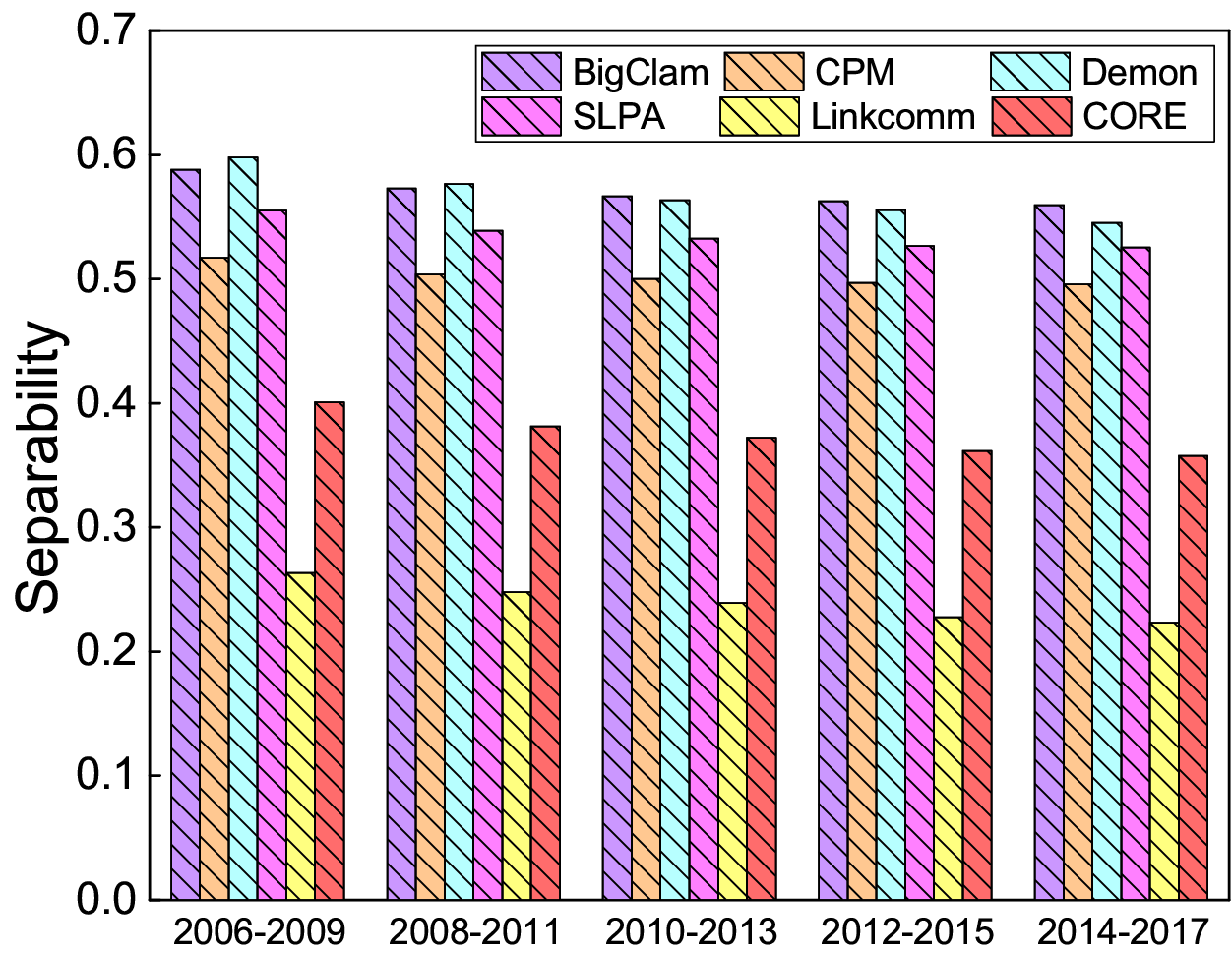}
    \label{fig:sep}
  }
  \subfigure[Average path length]{
    \includegraphics[width=0.33\textwidth]{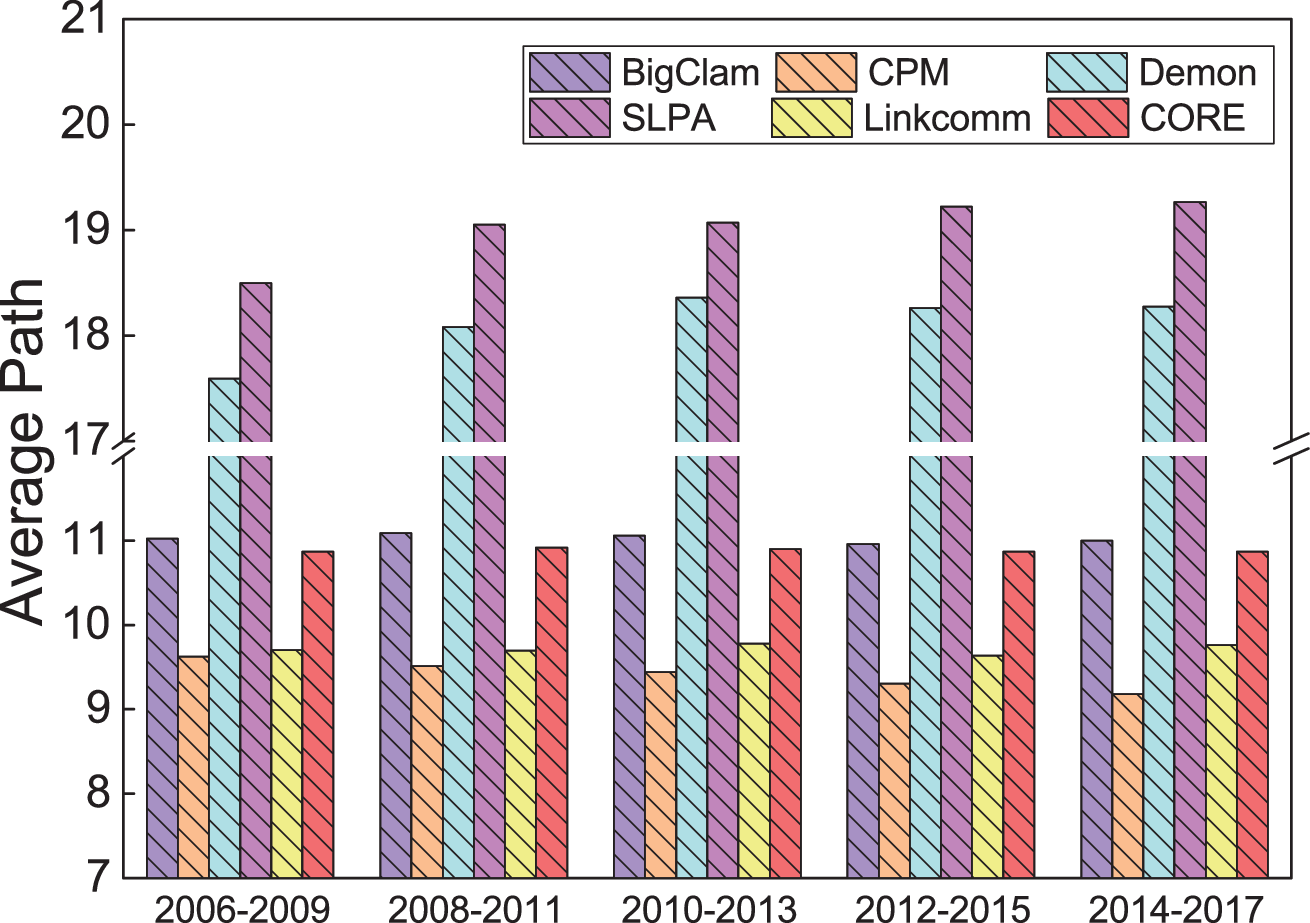}
    \label{fig:avgpath}
  }
  \caption{Comparison of different methods on average team size, separability, and average path length.}
\end{figure*}

\textbf{Average team size:} We then specifically analyzed the average numbers of team members, as shown in Figure~\ref{fig:avgnum}. As discussed above, Demon yielded the highest average number of team members, generally higher than 25. Linkcomm yielded the lowest average number of team members (approximately 5). CPM generated a relatively small average team size, whereas SLPA generated a relatively large average team size. Among these six methods, BigClam and CORE achieved the most reasonable average team sizes. Meanwhile, the average number of team members increased slowly over time.


\textbf{Separability:} Separability reflects the separation degrees between different teams. A higher Sep value refers to a loose team structure, which also indicates that the recognized team may not be close to reality. Figure~\ref{fig:sep} shows the distribution of Sep for the different methods during the 5 time periods. It can be seen that Linkcomm outperformed the rest, and our proposed CORE achieved the second-best position. The Sep values of the other 4 baseline methods (BigClam, CPM, Demon, and SLPA) were generally larger than 0.5. Therefore, we mainly discuss the performances of Linkcomm and CORE. According to Figure~\ref{fig:memdis}, the teams recognized by Linkcomm are generally extremely small, and it is easier for a smaller team to maintain a lower Sep than for a larger one.

We then analyzed the recognized CORE teams. There are over 40\% of edges connecting with external nodes, which is generally 10\%--30\% less than from Linkcomm and other baseline methods. In other words, our identified small-scale teams were better separated from the rest of the network than the others. Thus, teams of these sizes are manageable by their leaders.


\textbf{AvgPath:} Figure~\ref{fig:avgpath} shows the distribution of AvgPath. To better show the experimental results, we enlarged the value of AvgPath 10 times so that the differences between the results can be clearly recognized. AvgPath reflects tightness within recognized teams. It is one of the most significant network topology measures that can be used to verify the general distances between node pairs within teams.

As shown in Figure~\ref{fig:avgpath}, the AvgPath values of Demon and SLPA are considerably higher than those of the other methods. In combination with the team size statistics in Figure~\ref{fig:avgnum}, the teams recognized by Demon are generally larger. Therefore, teams recognized by Demon had higher AvgPath values. For SLPA, the high AvgPath value may be caused by the loose structures of the recognized teams.
\begin{figure*}[!tbp]
  \centering

  \subfigure[CCF]{
    \includegraphics[width=0.30\textwidth]{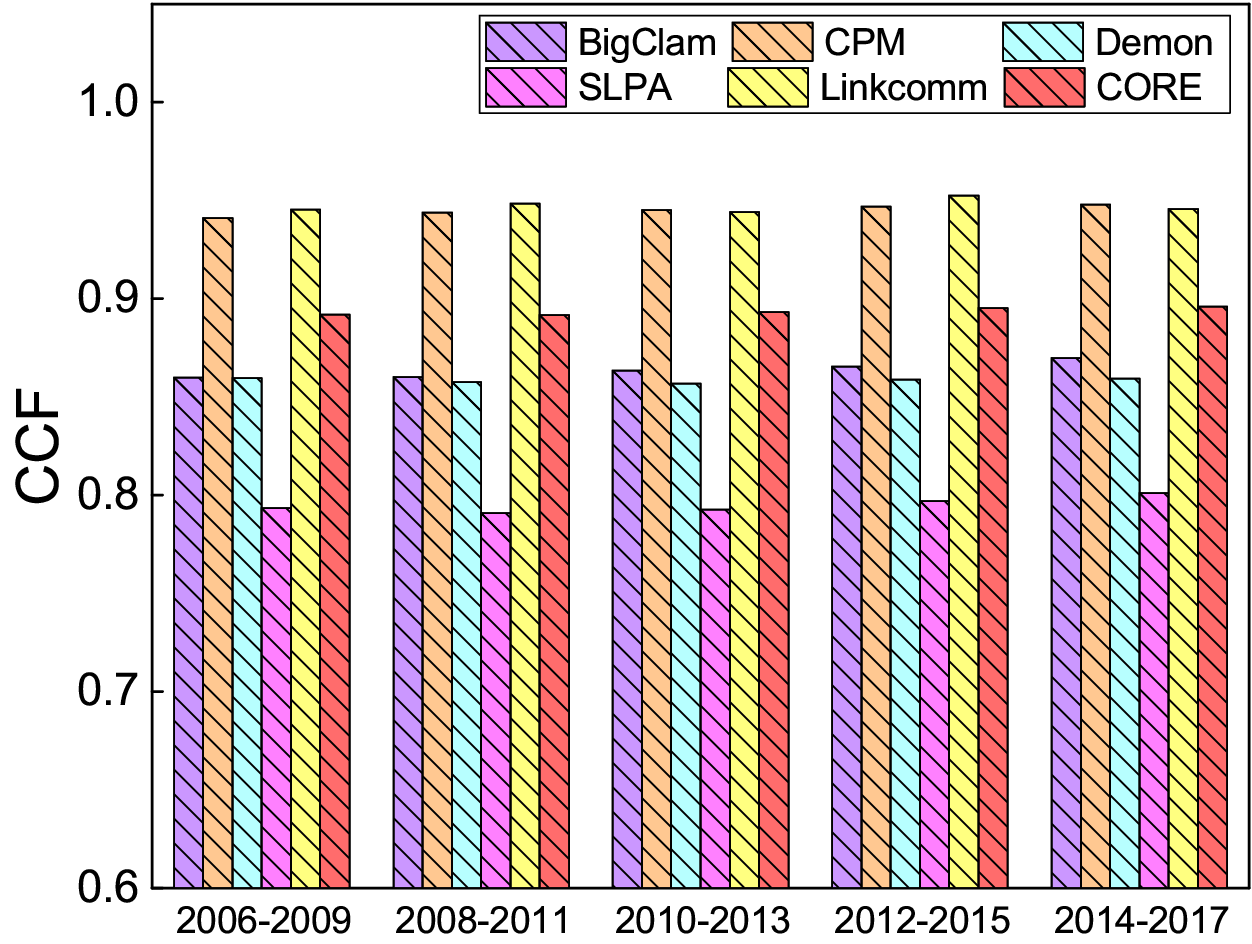}
    \label{fig:ccf}
  }
  \subfigure[ICDF]{
    \includegraphics[width=0.30\textwidth]{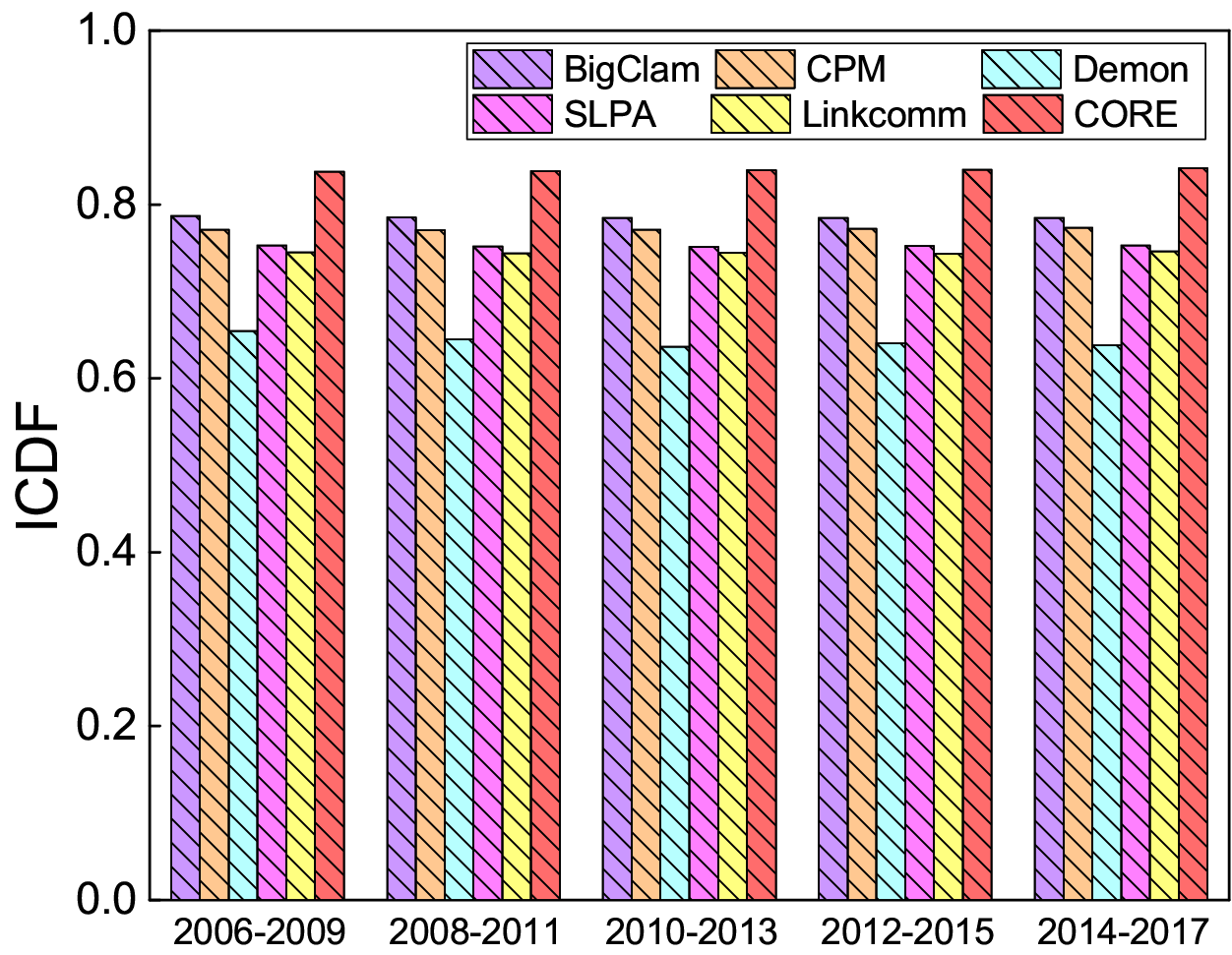}
    \label{fig:icdf}
  }
  \subfigure[Average number of triangles]{
    \includegraphics[width=0.30\textwidth]{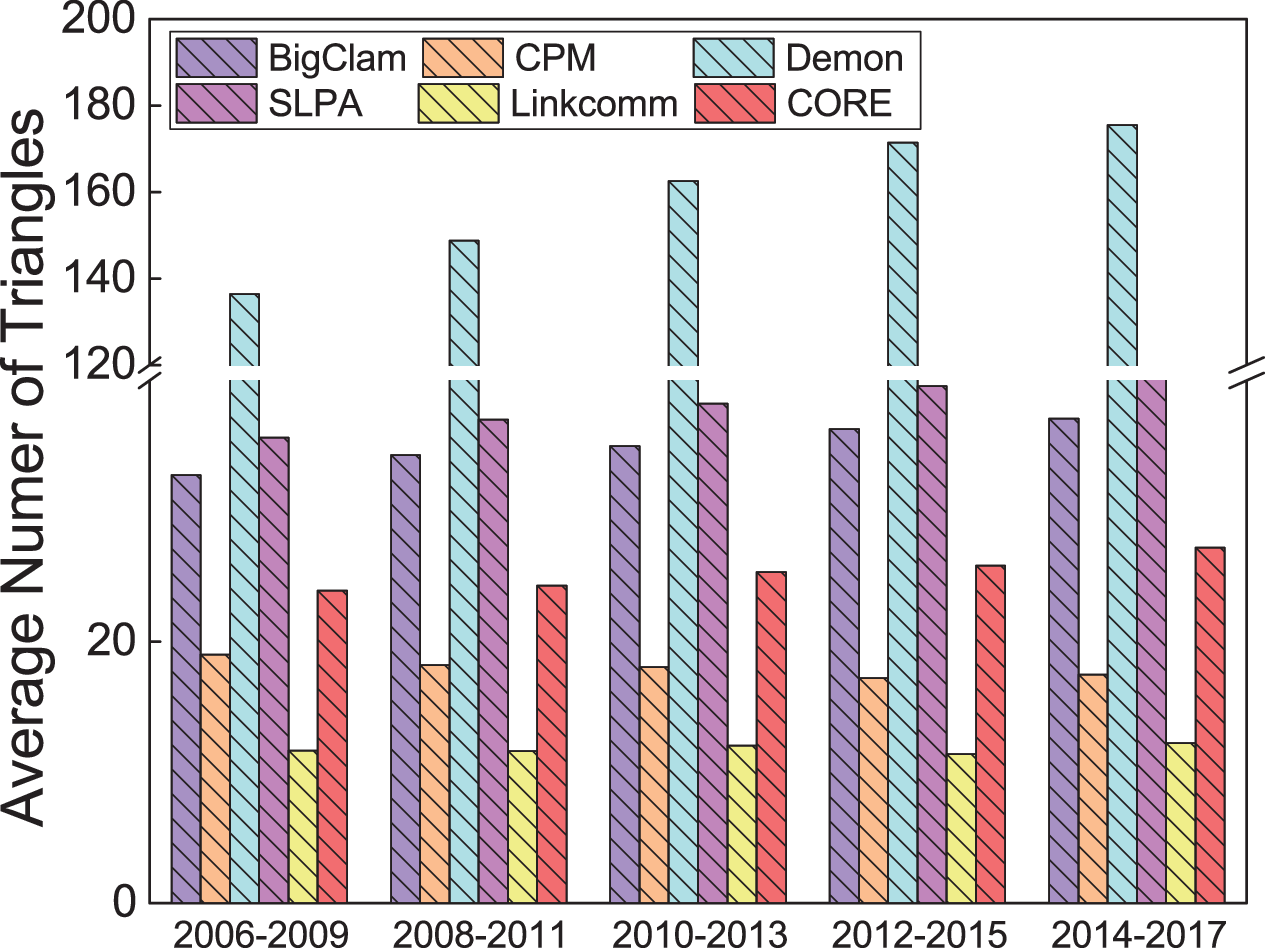}
    \label{fig:tri}
  }
\caption{Comparison of different methods on CCF, ICDF, and average number of triangles.}
\end{figure*}
Compared with Demon, SLPA, and BigClam, CORE has smaller AvgPath and closer teams, especially when team size increases. When the team size was larger than 35, the average path was approximately 19.4; thus, CORE outperformed the other algorithms by at least 10 percent. This indicates that the recognized teams have compact internal structures, meaning that team members have close relationships with each other. In comparison, Linkcomm and CPM performed better on this index because their overall team sizes were small. It is known that a lower value generally indicates better results for similar team sizes. Although the values of AvgPath of CORE are not as good as those of Linkcomm and CPM, they still maintain competitiveness on AvgPath in the team recognition task of the computer science discipline. This is because in computer science, the team size is generally between 6 and 7~\citep{fortunato2018science}. Different team sizes and structures will influence AvgPath to some extent. In this case, the metric should not be compared separately from the task. Therefore, for team recognition, especially in computer science, CORE achieves more accurate recognition results.

\textbf{CCF:} Figure~\ref{fig:ccf} shows the general trends of CCF for different algorithms. Linkcomm and CPM alternately occupy the first position, and CORE dominates the third position from the first year to the last year. As discussed above, Linkcomm recognizes extremely small teams, which makes CCF values relatively higher. Similar to Linkcomm, over 60\% of teams identified by CPM are of sizes in [3, 5]. Smaller team sizes usually lead to better CCF performance.

According to the statistics, CORE outperformed the other three methods by at least 7\% in the CCF. This also indicates that CORE achieves a closer team structure when the recognized team sizes are similar. When the team sizes were larger than 50, the performance of the proposed algorithm was found to be superior to that of other algorithms. However, the CCF values of the other algorithms were much lower than those of CORE, with declining trends. As mentioned above, the CCF is a commonly used metric for evaluating the internal structure of communities. Therefore, the proposed algorithm recognizes teams whose members are densely linked to locally inhomogeneous distributions.

\textbf{ICDF:} Figure~\ref{fig:icdf} shows the distribution of ICDF, in which CORE outperformed all other comparison algorithms. It can be seen that the ICDF of recognized teams by our proposed CORE is approximately 0.9 on average, meaning that there exists a member who collaborates with 90\% of their teams. However, other studies could not accurately detect this phenomenon. As illustrated above, a collaborative team with a higher ICDF indicates that there is a member connecting with more members. This also indicates that a collaborative team with a higher ICDF fits the proposed model, that is, the "core + extension" group model, better. The experimental results show that CORE can recognize collaborative teams closer to members with higher connectivity with other members. In other words, the other overlapping community detection algorithms cannot accurately recognize teams with "core + extension" structures.


\textbf{Triangles:} Figure~\ref{fig:tri} shows the average number of triangles within teams recognized by different methods. Compared with other methods, the CORE method has seemingly no advantages. However, the number of triangles is limited to the total number of nodes as well as the total number of edges. Therefore, we can explain the extremely high number of triangles in the teams recognized by Demon. This is because the teams recognized by Demon were generally large. Likewise, Linkcomm had the lowest average number of triangles because the recognized teams were too small.

\begin{figure}[!tp]
  \centering
  \includegraphics[width=0.6\textwidth]{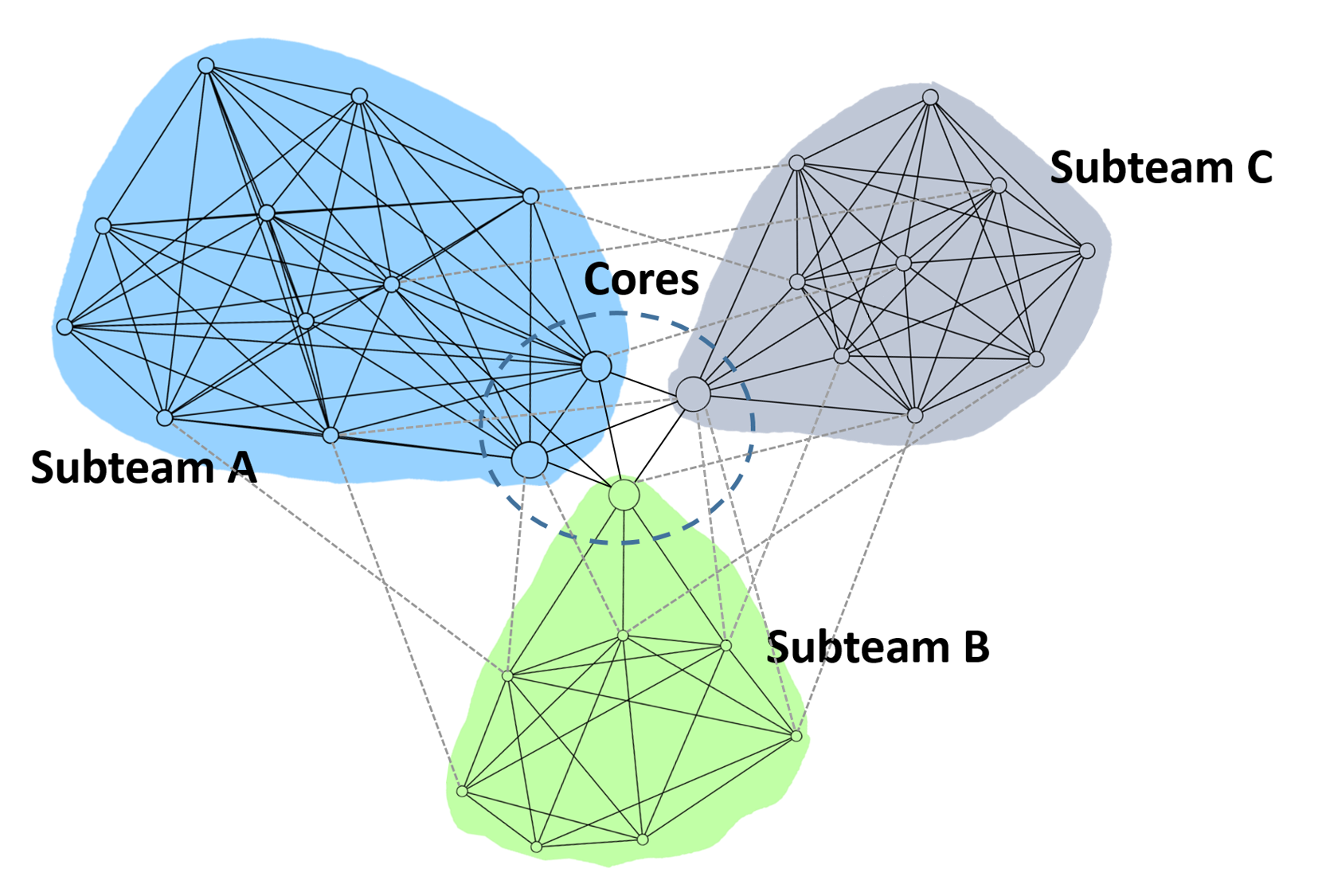}
  \caption{Sub-team structures in collaborative teams.}
  \label{fig:subteam}
\end{figure}

We then specifically analyzed the number of triangles when the number of team members was less than 40. We analyzed the team structures and found that most triangles existed between the core and extension members. However, extension members have few connections with different core members in the teams. We regard this type of structure in collaborative teams as a sub-team structure. In teams with a sub-team structure, core members in close relationships have several sub-teams consisting of their own affiliated extension group members. In particular, when teams grow larger, they often need a structure dominated by multiple cores to maintain the relationships of all members. The sub-team structure of collaborative teams is illustrated in Figure~\ref{fig:subteam}. Figure~\ref{fig:subteam} shows an extreme situation in which the three sub-teams do not collaborate despite the core members. In the real world, sub-teams typically have sparse connections with each other. As a typical type of social network, the recognition results of academic networks meet the characteristics of the six-degrees of separation theory (e.g., small-world theory). Moreover, the academic network is densely connected. This is because popular scholars attract more collaborators. Such "core + extension" phenomenon aligns with the natural property of social networks. Although members in different sub-teams rarely collaborate, they can easily communicate with each other through their cores. These collaborative behaviors lead to fewer triangles in our resulting teams compared with the other methods.

Summarizing the properties of our resulting teams, we found that CORE can identify collaborative teams more accurately in terms of team scale. According to the results in terms of Sep, AvgPath, and CCF, our method showed strong competitiveness in academic team recognition in computer science. Overall, the recognized teams have a more compact internal structure compared with other methods, indicating that team members maintain closer collaborative relationships with each other. In addition, the results of ICDF and triangles demonstrate the ``core + extension'' structure for teams, which also supports our preliminary definition of academic teams. In other words, academic teams are self-assembled. Self-assembly refers to the process in which components of a system organize into ordered or functional structures or patterns because of specific local interactions between the components themselves without external direction ~\citep{Grumezescu2016}. In academic networks, academic teams are self-assembled, meaning that a certain scholar is willing to join one academic team by themselves.
\subsection{Discussion}
Experimental results show that our proposed ``core + extension'' structure corresponds to scientific collaboration facts. CORE improves the recognition effectiveness and efficiency by using the high-order network topology attributes of members in the collaborative network. However, this is insufficient. In the real collaboration network, in addition to topological-level information, there are many features that are difficult to quantify or obtain, such as the relationship among team members and the working attitude. Importantly, these features often have a profound effect on the team 's collaboration output. How to measure these features qualitatively is of research value but challenge. In general, despite topology-level information, extracting semantic-level information and combining topology-level with it would be a feasible way to further improve the proposed CORE. Moreover, the computation complexity of CORE still needs to be reduced. In this paper, k-core decomposition method is employed as a rough way for identifying node importance. But when the search space grows large, k-core decomposition will become time-consuming and even NP-hard. This will lead to a high complexity of the algorithm for identifying core teams. Investigating more efficient distributed k-core decomposition algorithms may be a feasible way to reduce the computational complexity of the algorithm, which will be further practiced in follow-up studies.

\section{Exploring Collaboration Patterns}~\label{sec5}
Herein, we analyze collaboration patterns based on collaborative teams recognized by CORE.
\begin{table*}[htb]

  \renewcommand\arraystretch{1.3}
  \centering
  \caption{Comparison between all co-author networks and recognized collaborative team networks}
  \resizebox{\textwidth}{20mm}{
    \begin{tabular}{p{8.375em}cccccccccc}
      \toprule
      \multicolumn{1}{c}{Year} & \multicolumn{2}{c}{2006-2009} & \multicolumn{2}{c}{2008-2011} & \multicolumn{2}{c}{2010-2013} & \multicolumn{2}{c}{2012-2015} & \multicolumn{2}{c}{2014-2017} \\
      \hline
      \multicolumn{1}{c}{Network} & \multicolumn{1}{c}{Co-author} & \multicolumn{1}{c}{Team} & \multicolumn{1}{c}{Co-author} & \multicolumn{1}{c}{Team} & \multicolumn{1}{c}{Co-author} & \multicolumn{1}{c}{Team} & \multicolumn{1}{c}{Co-author} & \multicolumn{1}{c}{Team} & \multicolumn{1}{c}{Co-author} & \multicolumn{1}{c}{Team} \\
      \hline
      \multicolumn{1}{l}{Nodes} &      463,777  &      125,244  &              521,139  &            147,065  &              574,651  &            171,155  &              605,949  &            188,812  &              610,381  &            200,306  \\
      \multicolumn{1}{l}{Edges} &   1,254,015  &      479,612  &           1,445,722  &            584,526  &           1,630,940  &            698,943  &           1,767,594  &            793,566  &           1,837,752  &            856,476  \\
      \multicolumn{1}{l}{AvgDeg} & 5.408 & 7.659 & 5.548 & 7.949 & 5.676 & 8.167 & 5.834 & 8.406 & 6.022 & 8.552 \\
      \multicolumn{1}{l}{Components} &        43,390  &          2,005  &                48,589  &                2,128  &                53,685  &                2,419  &                54,616  &                2,674  &                51,100  &                2,787  \\
      \multicolumn{1}{l}{Triangles} &   1,496,758  &      685,593  &           1,757,927  &            847,718  &           2,024,557  &         1,033,864  &           2,244,693  &         1,200,434  &           2,399,702  &         1,319,204  \\
      CCF & 0.627 & 0.704 & 0.626 & 0.698 & 0.626 & 0.697 & 0.625 & 0.698 & 0.632 & 0.702 \\
      \bottomrule
    \end{tabular}%
  }
  \label{overall table}%
\end{table*}%
\subsection{Overall Review of the Collaboration Network}
To comprehensively understand collaborative patterns based on collaborative teams, we compared some basic properties, that is, nodes, edges, average degree (AvgDeg), graph-connected components, triangle numbers, and CCF, among all co-author networks and generated collaborative team networks in different time windows. In Table \ref{overall table}, collaborative team networks decline by nearly 70\% at the nodes and by 60\% at the edges, compared with the original co-author networks. However, these missing nodes and edges only lead to an approximately 45\% reduction in triangles, which is far less than the decreased ratios of nodes and edges. Meanwhile, we can also observe that the average degrees are approximately 8 in the collaborative team networks, meaning that every node has on average 8 neighbors in the networks. The average degree in the original co-author networks is only approximately 5.5, which is much lower than that of collaborative team networks. The CCF of the generated networks also increased by approximately 11\% compared with that of the co-author networks. In addition, the CORE algorithm can obtain a high-connectivity network based on the reduction in the number of graph-connected components. Changes in these basic properties demonstrate that the generated networks have densely connected structures. In other words, the proposed method successfully removed less important nodes and weak-tie edges.

\begin{figure}[ht]
  \centering
  \includegraphics[height=0.5\linewidth, width=0.7\linewidth]{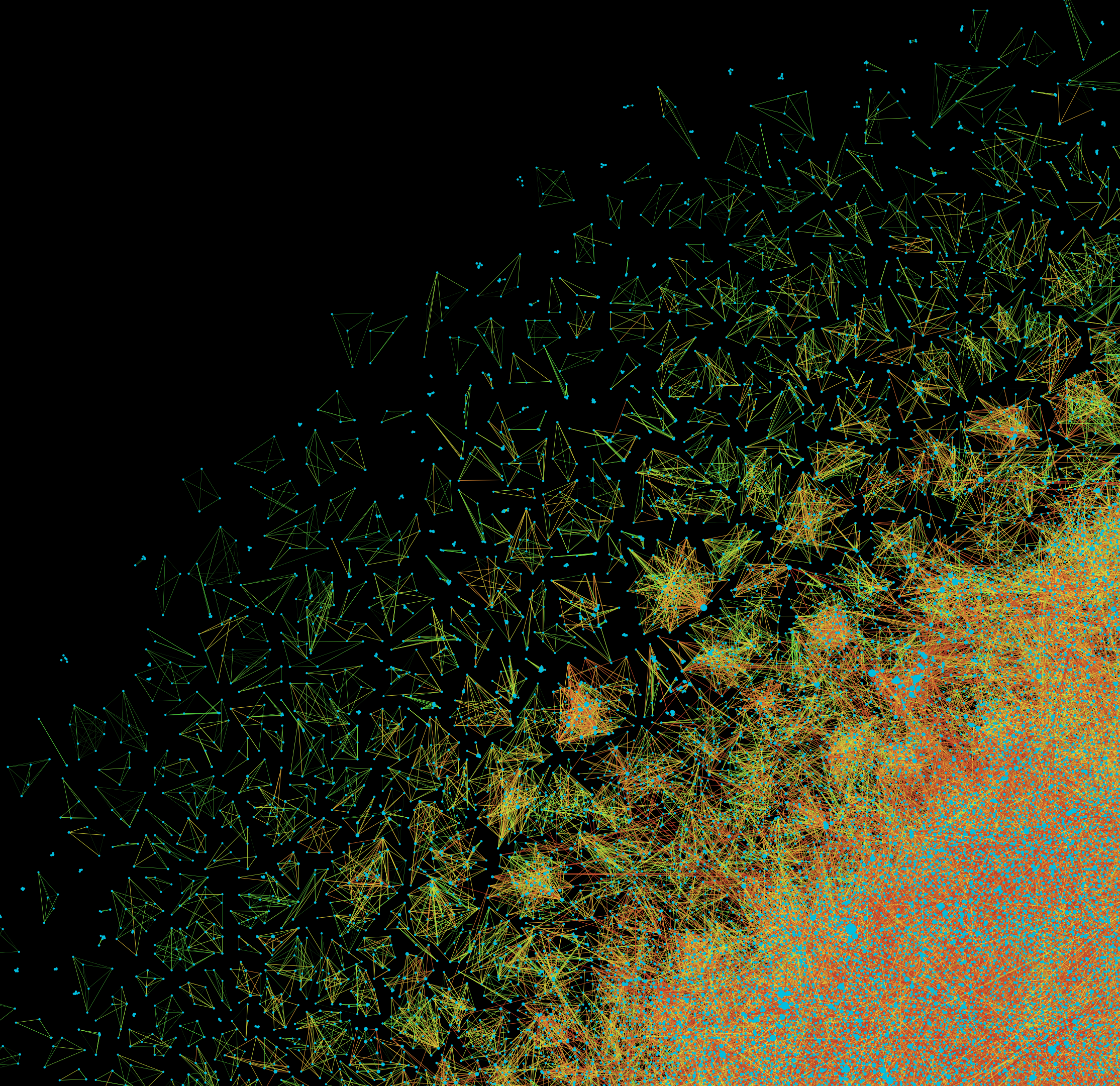}

  \caption{Overall collaboration network. In the figure, the blue nodes represent scholars, and edges with different colors are collaboration relationships with different CIIs. In addition, the bottom right corner is the center of the collaboration network, which is the maximally connected subgraph composed of scholars with high connectivity.}
  \label{overall1}
\end{figure}

In recognized collaborative team networks, nodes of different sizes represent scholars with different publication volumes, and the edges represent collaboration relationships between two scholars. Additionally, we arranged different labels for nodes to represent their teams and edges. Different colors (green-yellow-red in descending order) represent the values of CII, which reflect the closeness between scholars. Based on the analysis, we found that although collaboration networks in different time windows have different scales, they have similar structures. The generated collaboration network is shown in Figure \ref{overall1}. It can be seen from Figure \ref{overall1} that there is a trend for scholars to diverge from the center to the margin in the collaboration network. The center of the collaboration network is the maximally connected subgraph, which is composed of most scholars with high connectivity. However, the margin is composed of small-scale subgraphs that have few links with each other. 

\subsection{Analysis of Teams}
We then focused on the collaborative relationships among members of different collaborative teams. In Figure \ref{overall1}, we observe that teams pervading around the margin of networks mostly have green edges, which shows their close relationships with each team member. Center components of large scales generally have yellow or red edges, indicating that relationships are looser than those in the margins. In addition, by comparing the sizes of the nodes, it can be found that nodes with larger sizes are completely distributed in the center area. To explore these distributions quantitatively, we analyzed the team CII and performance. The team CII, which is also called the team average CII, is the average value of the CII of all connections in a team. In addition, we employed the team per-publication and per-citation numbers to evaluate teams' performance in different time windows.

\begin{itemize}
  \item The team per-publication number equals to all publications published by team members divided by the number of team members, which is calculated as
  \begin{equation}
    PP(t) = \frac{\sum_{i\in t}Pub(i)}{\left | t \right |},
  \end{equation}
  where $t$ is the set of team members, $ Pub(i) $ is the number of publications of scholar $ i $ during the time window, and $ \left | t \right | $ is the size of $ t $.

  \item The team per-citation number equals to all citations of papers published by team members divided by the number of team members, which is calculated as
  \begin{equation}
    PC(t) = \frac{\sum_{i\in t}Cit(i)}{\left | t \right |}.
  \end{equation}
  Here, $t$ is the set of team members, $ Cit(i) $ is the citation count of all articles of scholar $ i $ during the time window, and $ \left | t \right | $ is the size of $ t $.
\end{itemize}

These two metrics reflect the performance of a team in two different ways. A team with a large per-publication number is highly productive, and its members publish many papers in the time window. In addition, in a team with a large per-citation number, publications by team members are highly cited in their research fields, which partially reflects a team's research impact. Then, we analyzed variations in teams' average CII and team performance using these two metrics.

\begin{figure}[!t]

  \centering
  \subfigure[Per-publication number with CII of teams]{
    \includegraphics[width=0.4\linewidth]{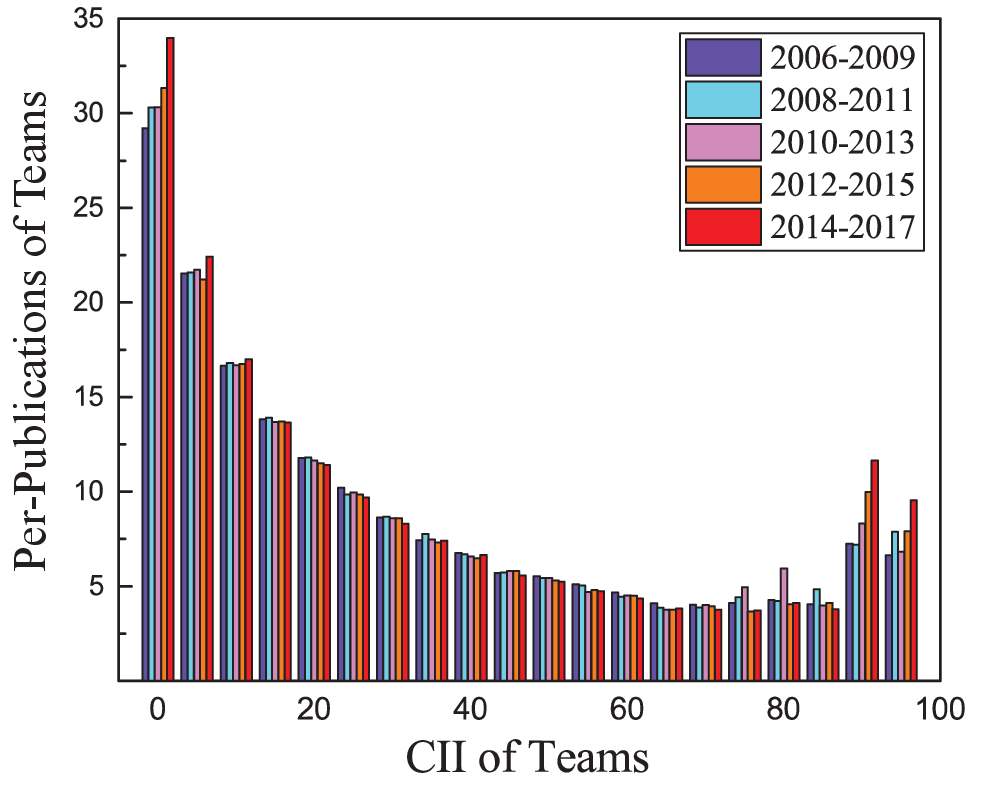}
    \label{fig:ciiper1}}
  \centering
  \subfigure[Per-citation number with CII of teams]{
    \includegraphics[width=0.41\linewidth]{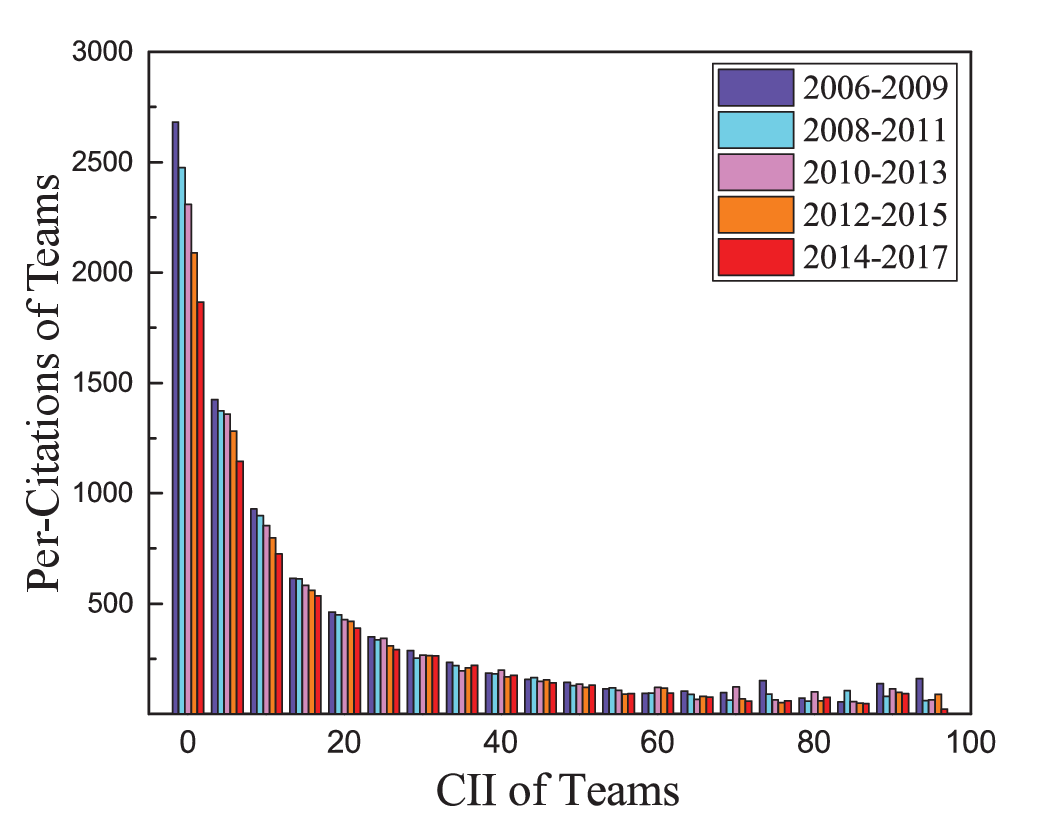}
    \label{fig:ciiper2}}
  \caption{Performance for core teams with different average CIIs in 5 time windows.}
  \label{fig:ciiperformances}
\end{figure}

In Figure~\ref{fig:ciiper1}, the horizontal axis represents the average CII of the collaborative team. That is, we first calculated the average CII of each recognized team and enlarged them 100 times because the original values of CII are generally small. We then categorized the enlarged CII values into different groups, with an interval of 5. Therefore, the first group refers to teams with an average CII in [0,5). The remainder of the process can be performed in the same manner. Figure~\ref{fig:ciiper1} shows the per-publication number of each team with different average CII values. It can be seen that the general distribution first descends with an increase in teams' average CII. When the teams' average CII is larger than 80, the team per-publication number rebounds. This is because of the way CII is defined: it can eliminate the influence of collaboration frequency. Simultaneously, a scholar with more collaborators will not have a higher average CII. This is because such scholars generally maintain multiple collaborations with different scholars. Thus, the CII is not an absolute metric but a relative metric that can be used to measure how closely two scholars collaborate.

There were some obvious distinctions between the different periods. When the CII was in [15, 70), most teams' per-publication numbers decreased slightly over time. However, this type of descent was not obvious. For the other CII segments, different year periods showed different performance. Generally, the team per-publication numbers grew over time, especially when the CII was extremely high or extremely low. Taking segment [0, 5) as an example, the team per-publication number in the year period of 2014-2017 was higher than in the other year periods.

A similar descending trend can be observed in Figure ~\ref{fig:ciiper2}. When the CII is relatively low, the team per-citation number is relatively high. It continues to decrease and increases slightly when the CII is in [70, 95).

\begin{figure}[!t]
  \centering
  \subfigure[Per-citation number with CII of core teams]{
    \includegraphics[width=0.42\linewidth]{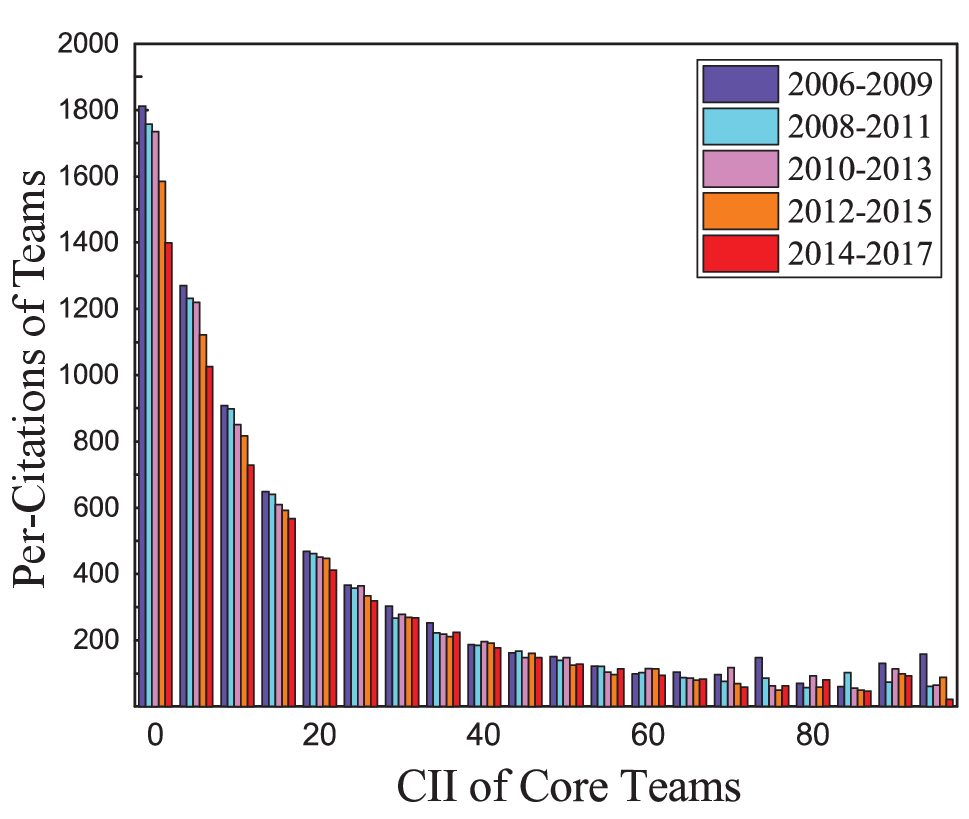}
    \label{fig:corecii}}
  \centering
  \subfigure[Average team h-index]{
    \includegraphics[width=0.4\linewidth]{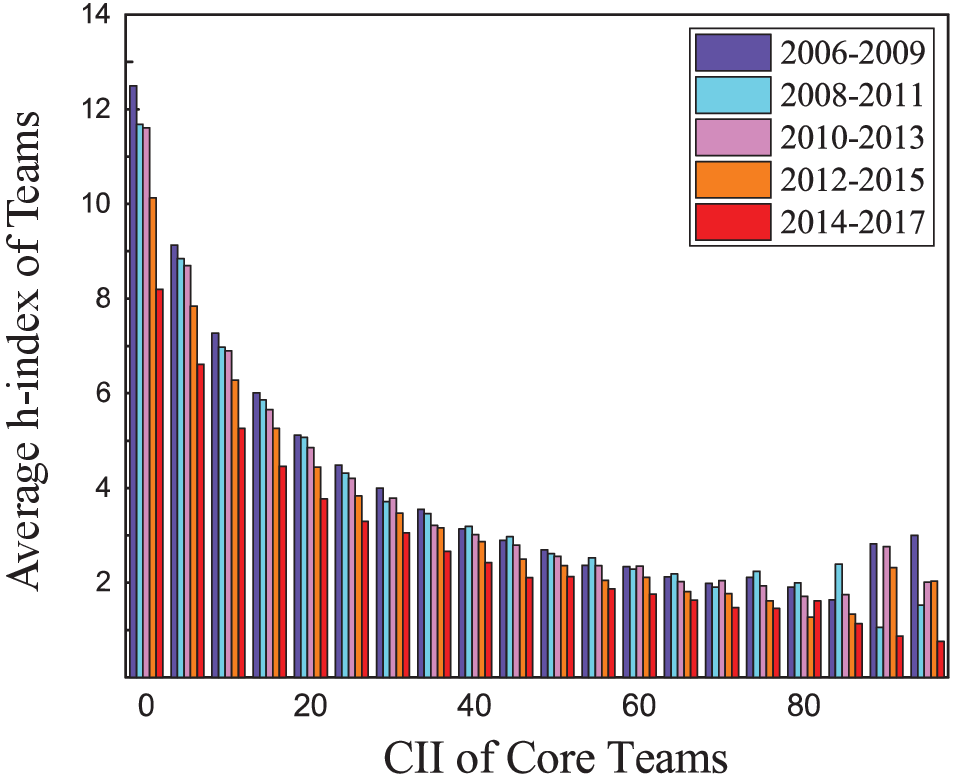}
    \label{fig:hcore}}
  \caption{Performance for teams with different average CIIs in 5 time windows.}
  \label{fig:coreperformance}
\end{figure}

We then analyzed the performance of the core members of the collaborative teams. Since core team members are those who occupy dominant positions in the team, we analyzed their abilities. A scholar with a higher h-index value often indicates that they gained a great reputation in science. Therefore, we specifically analyzed two perspectives for core teams: per-citation number and average h-index.

Figure~\ref{fig:corecii} indicates that core teams show a similar trend to regular teams, demonstrating a clear descending trend. Moreover, when the CII is in [70,90), the per-citation number fluctuates with a general increasing trend. A researcher's h-index refers to the fact that at most h of their papers are cited by other papers at least h times. Figure~\ref{fig:hcore} shows the h-index distribution with the increasing average CII of the team: scholars stay in teams with tight collaborative relationships if they have low h-indices. These dynamic changes indicate that scholars with low h-indices often have stable collaborators, that is, their team members. Scholars with more publication records not only maintain low-intensity collaboration relationships with their team members but also seek a number of diverse collaborators to accomplish their various outputs.

Figure ~\ref{fig:ciiperformances} shows the performance of teams with an increasing team average CII. Figure \ref{fig:coreperformance} shows that teams with a lower CII are highly productive in all time windows, and the production of teams has a clear declining trend with increasing average CII. However, when the average team CII is higher than 85, the per-publication number increases slightly. By analyzing these teams and their publications, we find that less than 1\% of all teams have an average CII of more than 85. In addition, by observing the collaborative networks of these core teams, the CII values between nodes in the team are much higher than the weight of edges between nodes outside the team, which shows that they barely cooperate with anyone except their team members, and their papers often have few citations. It can also be seen from Figure \ref{fig:ciiper2} that the quality of the teams also has a clear declining trend with increasing CII in all time windows. These two figures show that teams with a lower CII have a better performance. Teams with members with broad and low-intensity collaborative relationships have excellent performance.

\begin{figure}[htbp]
  \centering

  \subfigure[2008]{
    \includegraphics[width=0.22\textwidth]{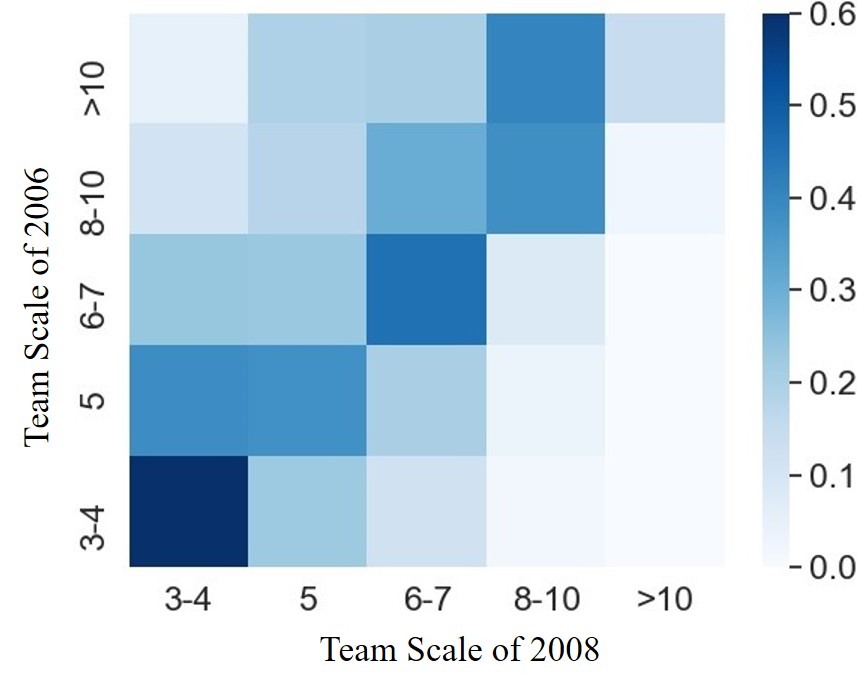}
  }
  \subfigure[2010]{
    \includegraphics[width=0.22\textwidth]{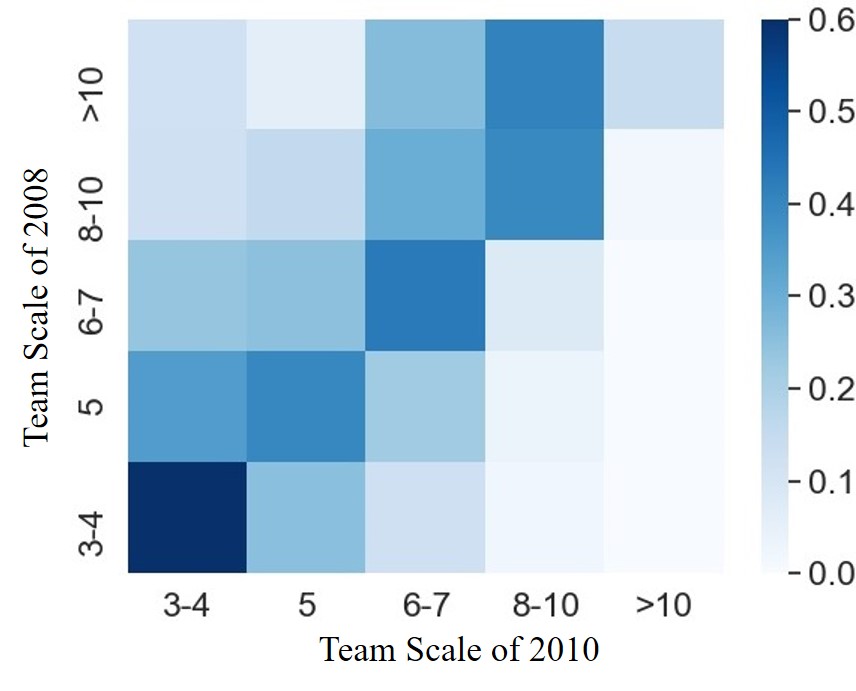}
  }
  \subfigure[2012]{
    \includegraphics[width=0.22\textwidth]{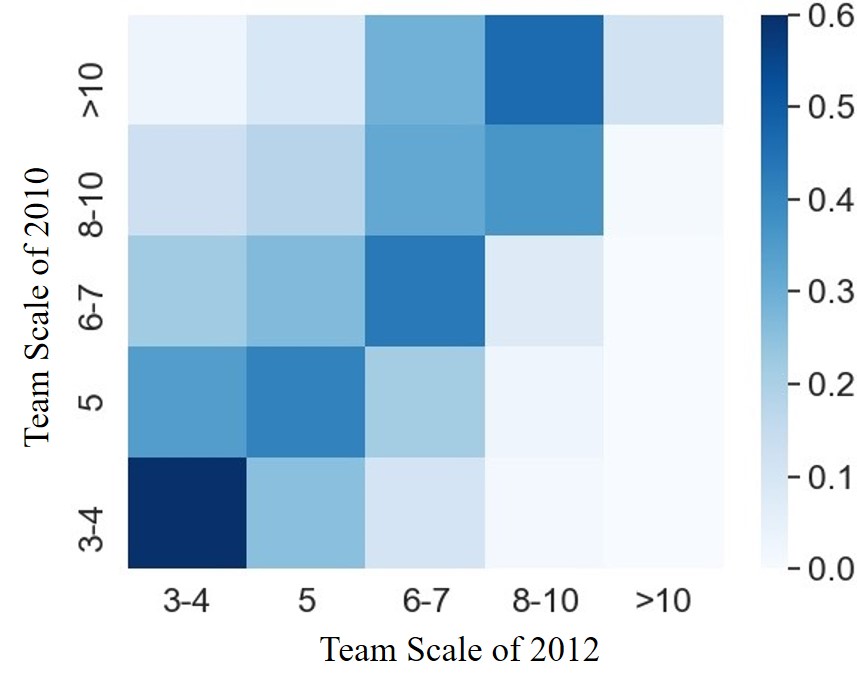}
  }
  \subfigure[2014]{
    \includegraphics[width=0.22\textwidth]{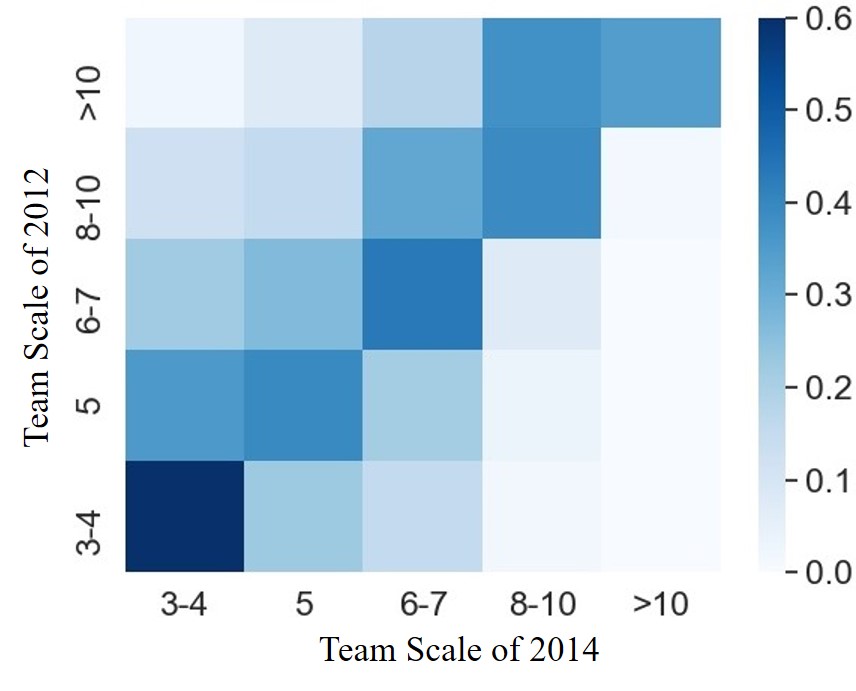}
  }
  \caption{Statistics of turnover rates in academic teams.}
  \label{fig:flow}
\end{figure}

Furthermore, it is also of great interest to explore patterns in the academic career movement. This flow is represented by counting the number of scholars who left their teams in a certain year. Here, we use the turnover rate as an indicator to measure team flow, which can be calculated as
\begin{equation}
    Turnover = \frac{y_{i,j}}{y_{total,i}},
\end{equation}
where $y_{i,j}$ represents the number of scholars transferred from team of size $i$ to team of size $j$ this year, and $y_{total,i}$ represents the total number of scholars transferred from team of size $i$ to other teams. For example, $y_{5,6}$ is the number of scholars who transferred from team of size 5 to team of size 6, and $y_{total,5}$ is the total number of scholars transferred to other teams from a team of size 5. In our study, we chose the time windows that we have already used to calculate the turnover rate (i.e., 2008-2011, 2010-2013, 2012-2015, and 2014-2017). We used our recognized teams as a basis and determined the turnover rate of teams at different scales (i.e., 3--4, 5, 6--7, 8--10, more than 10 members). Turnover rates were calculated among distinct team scales. The corresponding results are shown in Figure~\ref{fig:flow}. It can be seen that academic teams generally have high turnover rate, wherein the mobility of smaller teams is much higher than that of larger teams in all years. It appears that the main turnover rate distributions were similar, but the turnover rate of large teams increased significantly in 2014. Formulating academic teams to enhance scientific collaboration has become a fundamental pattern in academia. Stable collaborative teamwork brings about output of higher quality~\citep{Rand17093}, which may lead to a growing team formation rate. Meanwhile, the cost of career moves in academia also decreases owing to the development of transportation as well as information technologies. In addition, complicated scientific problems require the involvement of a larger academic team. Therefore, multiple underlying reasons may have an integrated function in this phenomenon.

\begin{figure*}[ht]
  \centering

  \subfigure[2008]{
    \includegraphics[width=0.44\textwidth]{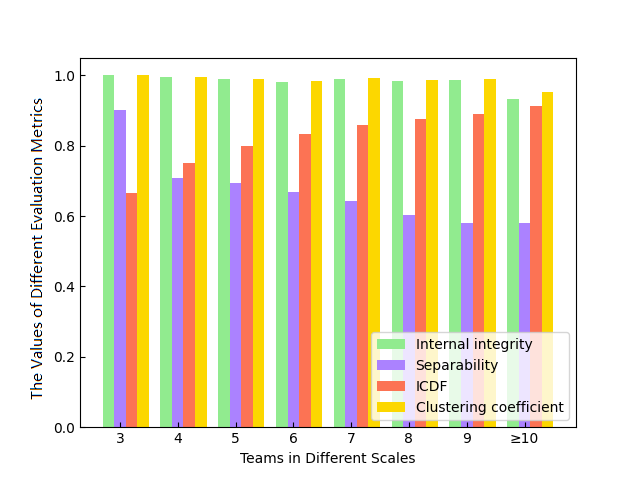}
  }
  \subfigure[2010]{
    \includegraphics[width=0.44\textwidth]{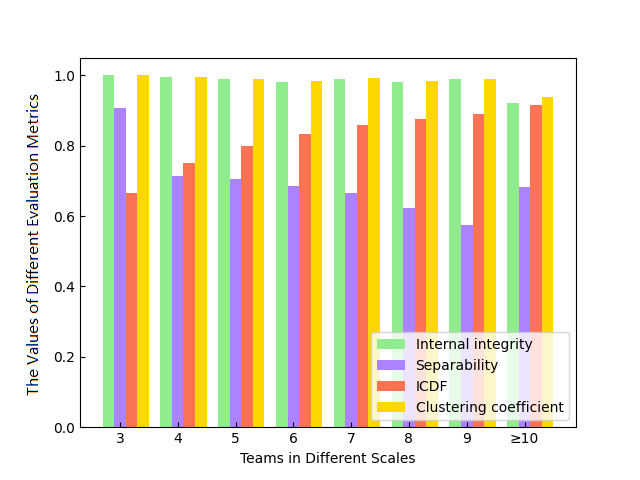}
  }\\
  \subfigure[2012]{
    \includegraphics[width=0.44\textwidth]{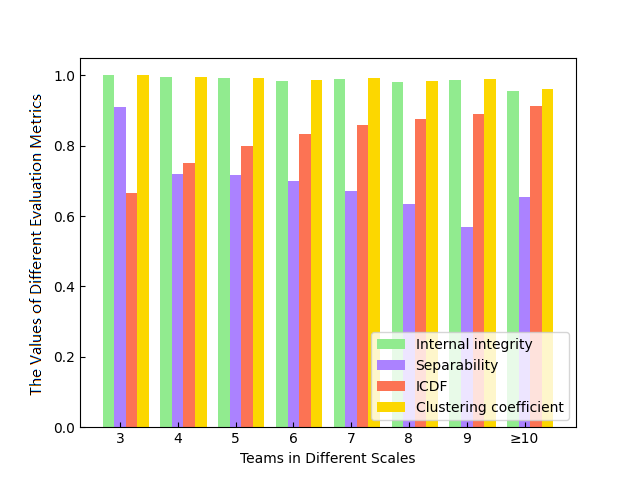}
  }
  \subfigure[2014]{
    \includegraphics[width=0.44\textwidth]{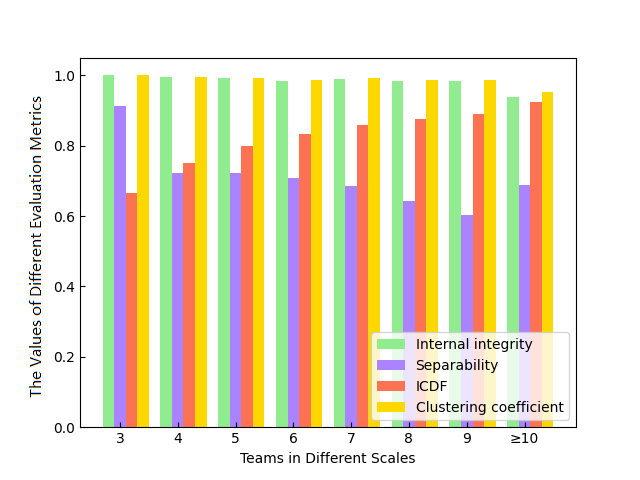}
  }
  \caption{Statistics of internal integrity, separability, ICDF, and clustering coefficient for teams in different scales.}
  \label{fig:kcore}
\end{figure*}

Four metrics (internal integrity, separability, ICDF, and clustering coefficient) were analyzed for different year groups, as shown in Figure~\ref{fig:kcore}. Internal integrity is an indicator that measures the gap between a graph and a complete graph and is calculated as 
\begin{equation}
    II = \frac{2*e}{l*(l-1)}.
\end{equation}
Here, $e$ is the number of edges, and $l$ is the number of nodes. In other words, internal integrity refers to the ratio of the edges in the subgraph to its corresponding complete graph. All statistical information was generated from teams recognized by the CORE. The main trends of internal integrity, ICDF, and CCF in the four years were quite close, whereas Sep showed a distinct distribution. Large teams have lower internal integrity and CCF than smaller teams. The ICDF increased with the scale of the team. Sep is expected to decrease with the growth of team scale. However, except for 2008, all teams with more than 10 members in the other three years (i.e., 2010, 2012, and 2014) had higher Sep than smaller teams. This phenomenon has shown interesting patterns in academia. Smaller teams have cohesive collaborative relationships. Consequently, Sep in smaller teams is generally higher. With the growing scale of academic teams, collaborative relationships are becoming sparser, which is the main reason for the decrease in Sep. However, teams with 10 or more members having a higher Sep may be caused by many reasons. For example, a large team project will maintain its structure for a period of time, but it will also naturally separate into several teams when the project is completed. This results in a greater degree of separation, leading to a higher Sep. In large teams, there are many repeated collaborations among multiple members, which will also lead to a higher Sep. It was found that the number of large teams increased rapidly, which was verified by our statistical results~\citep{wu2019large}. Larger teams may have frequent collaborations, thus leading to more output, which can be reflected significantly in co-author networks. Because the proposed CORE is based on such networks, the Sep of large teams may be higher than that of other teams.

Based on the above analysis, we conclude that teams with low-intensity collaborative relationships perform better than teams consisting of scholars who primarily complete publications with their team members. Furthermore, scholars with high reputations and mature careers prefer to maintain their relationships not only with their team members but also with other collaborators. These conclusions explain why scholars have gradually broadened their collaborative networks. This microscopic collaborative mechanism (i.e., collaborative patterns between nodes) motivates scholars to pursue more collaborators with higher levels, which results in widespread collaborations in scientific models from a macroscopic perspective ~\citep{fortunato2018science,azoulay2018toward,rogerTeam2005}.

\begin{table}[ht]
  \caption{Performance of core members and teams in different time windows}
  \begin{tabular}{cccccc}
    \hline
    & 2006-2009 & 2008-2011 & 2010-2013 & 2012-2015 & 2014-2017 \\ \hline
    Epub & 7.6       & 7.7       & 7.5       & 7.5       & 7.3       \\
    Cpub & 16.3      & 16.6      & 16.4      & 16.3      & 16.6      \\
    Ecit & 337.5     & 323.0     & 297.5     & 272.4     & 235.5     \\
    Ccit & 892.4     & 866.3     & 842.1     & 788.3     & 725.6     \\
    EAA  & 6.1       & 6.3       & 6.6       & 6.8       & 7.0       \\
    CAA  & 9.9       & 10.1      & 10.5      & 10.8      & 11.2      \\ \hline
  \end{tabular}
  \label{Core_team}
\end{table}

As we constructed teams with the "core + extension" composition mechanisms, they guaranteed the structural superiority of core members in their own teams. We then analyzed the performances of cores and their team extensions mainly from 3 perspectives. Table \ref{Core_team} reports average numbers of publications of extensions (Epub) and core members (Cpub), average numbers of citations of extensions(Ecit) and core members (Ccit), and average academic ages of extensions (EAA) and core members (CAA) in different time windows.

We noticed that Cpubs are greater than Epubs, which indicates that the core members publish approximately 9 articles on average more than the extension members. In addition, Ccits are much greater than Ecits, which shows that the core members' publications have a higher quality compared with those of the extension members. We also observed that the academic ages of extension members are lower than those of core members by about 4 years. Therefore, core members are more mature and perform better than extensions. These superiorities in performance and experience result in the leadership of core members in their own teams, which attracts more scholars to join in order to broaden their collaboration networks.

\section{Conclusion}~\label{sec6}

	In this study, we investigated collaboration patterns among scholars by recognizing collaborative teams in computer science from a large co-author network constructed using the MAG-CS dataset. The proposed CORE method takes account of "core + extension" structure to recognize collaborative teams, which firstly evaluated collaborative relationships quantitatively with a new index CII. Second, we used a series of scientific evaluation metrics and structural properties of networks to identify core team members. Then, we recognized the extensions attached to the cores by defining the leadership distribution function and optimizing the influential entropy of the network. We evaluated the performance of CORE using the proposed metrics for collaborative teams. Compared with state-of-the-art methods, the proposed algorithm demonstrated promising results.
	
	It is recognized that exploring collaborative patterns of academic teams can provide better services for academic society. This study may provide guidance for the development of academic team construction and evaluation. Since extensive investigation of academic teams is limited by the lack of real-world academic team datasets, this work can provide a feasible solution for automatically recognizing academic teams. As teams become the basic unit in science discovery, our empirical study reveals a new aspect of mining collaboration patterns, which is based on collaborative behaviors of different teams. The discovered phenomena align with the facts of scientific collaboration. This study clarifies team building, which determines the structure of collaborative networks and team performance. In this study, we focus only on the evaluation of the topology level in academic teams. However, semantic-level information also occupies a significant position in the investigation of team recognition. It may be feasible to evaluate and improve methods through both semantic and topological levels. Furthermore, we pioneered a new line of study to understand collaborative networks by analyzing collaboration teams at a mesoscopic level. These findings are expected to lead to a comprehensive understanding of collaborative patterns in academia.
	
\section*{Acknowledgments}
This work was funded by King Saud University, Riyadh, Saudi Arabia under the Researchers Supporting Project No. RSP2022R509 and National Natural Science Foundation of China under Grants No. 61872054 and 62102060. The authors thank Huan Liu (Arizona State University), Hanghang Tong (UIUC), Shirui Pan (Monash University), Kaiyuan Zhang, Yuchen Sun, and Xiangtai Chen (Dalian University of Technology) for their help with this work.




\bibliographystyle{cas-model2-names}

\bibliography{ref}
\end{document}